\documentclass[nologos]{ieeeaccess}
\usepackage{amsmath,amssymb,amsfonts}
\usepackage{algorithmic}
\usepackage{graphicx}
\usepackage{textcomp}
\usepackage{algorithm}
\usepackage{booktabs}
\usepackage{physics}
\usepackage{algorithm}
\usepackage{listings}
\usepackage[table,dvipsnames]{xcolor}
\usepackage{pict2e}
\usepackage{silence}
\lstdefinestyle{rust}{
    basicstyle=\ttfamily\scriptsize\mdseries,
    keywordstyle=\color{orange!75!red}\bfseries,
    commentstyle=\color{gray!50!black}\itshape,
    stringstyle=\color{red!60!black},
    identifierstyle=\color{black},
    breaklines=true,
    breakatwhitespace=true,
    frame=leftline,
    framerule=1.5pt,
    rulecolor=\color{orange!50!red},
    backgroundcolor=\color{gray!3},
    tabsize=2,
    showstringspaces=false,
    xleftmargin=12pt,
    xrightmargin=4pt,
    framexleftmargin=8pt,
    comment=[l]//,
    morecomment=[s]{/*}{*/},
    morekeywords={let, mut, move, fn, impl, struct, enum, trait, pub, use, usize, f64, i32, i64, bool, String, Vec, Option, Result, Circuit, ParameterizedCircuit, Circuit::new, params},
    keywordsprefix=,
    aboveskip=6pt,
    belowskip=6pt
}
\lstdefinestyle{python}{
    language=Python,
    basicstyle=\ttfamily\scriptsize\mdseries,
    keywordstyle=\color{blue!70!violet}\bfseries,
    commentstyle=\color{green!50!black}\itshape,
    stringstyle=\color{purple!70!black},
    identifierstyle=\color{black},
    breaklines=true,
    breakatwhitespace=true,
    frame=leftline,
    framerule=1.5pt,
    rulecolor=\color{blue!60!violet},
    backgroundcolor=\color{blue!2},
    tabsize=4,
    showstringspaces=false,
    xleftmargin=12pt,
    xrightmargin=4pt,
    framexleftmargin=8pt,
    morekeywords={def, return, True, False, None, qml},
    aboveskip=6pt,
    belowskip=6pt
}
\PassOptionsToPackage{colorlinks=false,pdfborder={0 0 0}}{hyperref}
\newsavebox{\ORCIDlogo}
\savebox{\ORCIDlogo}{%
  \setlength{\unitlength}{\dimexpr 1em/256\relax}%
  \begin{picture}(256,256)%
    \color[HTML]{A6CE39}\put(128,128){\circle*{256}}%
    \color{white}%
    \put(78.6,199.2){\circle*{20}}%
    \moveto(70.9,176.9)\lineto(86.3,176.9)\lineto(86.3,69.8)\lineto(70.9,69.8)%
    \closepath\fillpath%
    \moveto(108.9,176.9)\lineto(150.5,176.9)%
    \curveto(190.1,176.9)(207.5,148.6)(207.5,123.3)%
    \curveto(207.5,95.8)(186,69.7)(150.7,69.7)%
    \lineto(108.9,69.7)%
    \closepath\fillpath%
    \color[HTML]{A6CE39}%
    \moveto(124.3,83.6)\lineto(148.8,83.6)%
    \curveto(183.7,83.6)(191.7,110.1)(191.7,123.3)%
    \curveto(191.7,144.8)(178,163)(148,163)%
    \lineto(124.3,163)%
    \closepath\fillpath%
  \end{picture}%
}
\newcommand{\orcidlink}[1]{%
  \href{https://orcid.org/#1}{\raisebox{-0.1ex}{\usebox{\ORCIDlogo}}}%
}
\usepackage{hyperref}
\usepackage[%
  backend = biber,%
  maxbibnames = 10,%
  minbibnames = 2,%
  style = ieee,%
  citestyle = numeric-comp,%
  bibencoding = utf8,%
  datamodel = software,%
  defernumbers = true,%
  sortcites = true,%
  doi = true,%
  isbn = false,%
  url = false,%
  eprint = true%
]{biblatex}
\DeclareFieldFormat{journaltitle}{\textit{#1}}
\renewbibmacro*{doi+eprint+url}{%
  \printfield{doi}%
  \newunit\newblock
  \printfield{eprint}}

\def\BibTeX{{\rm B\kern-.05em{\sc i\kern-.025em b}\kern-.08em
    T\kern-.1667em\lower.7ex\hbox{E}\kern-.125emX}}

\newdimen\xfigwd

\begin{document}
\history{Date of publication xxxx 00, 0000, date of current version xxxx 00, 0000.}
\doi{10.1109/TQE.2020.DOI}
\title{LogosQ: A High-Performance and Type-Safe Quantum Computing Library in Rust}
\author{\uppercase{Shiwen An}\authorrefmark{1}\orcidlink{0000-0002-4727-2906},
\uppercase{Jiayi Wang}\authorrefmark{2}, and \uppercase{Konstantinos Slavakis}\authorrefmark{1}\orcidlink{0000-0002-3370-3154}}
\address[1]{School of Engineering, Institute of Science Tokyo, Yokohama, Kanagawa, 226-8502 Japan}
\address[2]{School of Computer Science, Georgia Institute of Technology, Atlanta, GA, 30332 USA}
\tfootnote{This work is supported by Super Smart Society (SSS) Project of the Ministry of Education, Culture, Sports, Science and Technology (MEXT), Japan and Spring Fellowship at Institute of Science Tokyo.}

\markboth
{Shiwen An \headeretal: Preparation of Papers for IEEE Transactions on Quantum Engineering}
{Konstantinos Slavakis \headeretal: Preparation of Papers for IEEE Transactions on Quantum Engineering}

\corresp{Corresponding author: Shiwen An, email: an.s.aa@m.titech.ac.jp}

\begin{abstract}
Developing robust and high-performance quantum software is challenging due to the dynamic nature of existing Python-based frameworks, which often suffer from runtime errors and scalability bottlenecks. In this work, we present LogosQ, a high-performance backend-agnostic quantum computing library implemented in Rust that enforces correctness through compile-time type safety. Unlike existing tools, LogosQ leverages Rust's static analysis to eliminate entire classes of runtime errors, particularly in parameter-shift rule gradient computations for variational algorithms. We introduce novel optimization techniques, including direct state-vector manipulation, adaptive parallel processing, and an FFT-optimized Quantum Fourier Transform, which collectively deliver speedups of up to 900$\times$ for state preparation (QFT) and 2--5$\times$ for variational workloads over Python frameworks (PennyLane, Qiskit), 6--22$\times$ over Julia implementations (Yao.jl), and competitive performance with Q\#. Beyond performance, we validate numerical stability through variational quantum eigensolver (VQE) experiments on molecular hydrogen and XYZ Heisenberg models, achieving chemical accuracy even in edge cases where other libraries fail. By combining the safety of systems programming with advanced circuit optimization, LogosQ establishes a new standard for reliable and efficient quantum simulation.
\end{abstract}

\begin{keywords}
Quantum computing, quantum simulation, variational quantum algorithms, parameter-shift rule, compile-time type safety, quantum circuits, matrix product states, Rust programming language.
\end{keywords}

\titlepgskip=-15pt

\maketitle

\section{Introduction}
\label{sec:introduction}

Quantum software has become the foundation of the quantum computing ecosystem, providing essential tools and libraries for developing, simulating, and executing quantum algorithms on various quantum hardware platforms. The field encompasses diverse frameworks tailored to different needs: Python-based libraries such as PennyLane \cite{bergholm_pennylane_2022}, which enables seamless integration of quantum circuits with classical machine learning models for variational algorithms \cite{cerezo_variational_2021}; Qiskit \cite{javadi-abhari_quantum_2024}, providing comprehensive tools for quantum circuit design and execution on IBM's quantum processors; and Cirq \cite{cirq_developers_cirq_2025}, focusing on near-term quantum devices \cite{preskill_quantum_2018} with frameworks tailored to Google's hardware. Julia-based frameworks include Yao.jl \cite{luo_yaojl_2020}, emphasizing parallel computing and CUDA optimization. Lower-level tools include Quantum++ \cite{gheorghiu_quantum_2018}, a C++11 library focusing on performance and portability, and OpenQASM \cite{cross_openqasm_2022}, a standardized assembly language promoting interoperability across quantum platforms. For variational quantum algorithms, matrix product state (MPS) based simulators have emerged as a scalable alternative to state-vector methods, enabling simulation of larger quantum chemistry systems through tensor network representations \cite{guo_differentiable_2023}. Recent advances in differentiable programming with tensor networks \cite{liao_differentiable_2019} have enabled automatic differentiation of quantum circuits and tensor network algorithms, providing efficient gradient computation for variational optimization. Additionally, automatically differentiable quantum circuits have been developed for many-qubit state preparation using MPS representations \cite{zhou_automatically_2021}. Recent research has addressed practical challenges through scalable benchmarks \cite{hines_scalable_2024}, mutation testing of quantum programs \cite{fortunato_mutation_2022}, and real-world applications such as hash function pre-image attacks \cite{preston_applying_2022}. Ongoing efforts in large-scale quantum simulators \cite{ismail_transversal_2025} and parameterized quantum circuit generators \cite{mao_q-gen_2025} continue to expand the field's capabilities.

The emergence of Rust has catalyzed new quantum simulation tools, with Spinoza \cite{yusufov_designing_2023} demonstrating Rust's potential for high-performance state-vector simulation through optimized amplitude pair manipulation. However, Spinoza focuses primarily on fast gate-level state vector simulation and does not address the broader needs of variational quantum algorithms, including parameterized circuit management, gradient computation, and scalable tensor network backends. Despite these advances, existing quantum software libraries face persistent challenges in extensibility, performance, and correctness---particularly silent runtime errors in gradient computations for variational algorithms. 

\subsection{Contributions}
To address these challenges, we present LogosQ, an extensible circuit-oriented quantum computing library implemented in Rust, a systems programming language known for its speed, memory efficiency, and thread safety \cite{matsakis_rust_2014}. Unlike Spinoza, which is primarily a state vector simulator, LogosQ is a comprehensive quantum computing library that prioritizes compile-time type safety for variational algorithms while combining performance, safety, and ease of use. The key differentiators from existing Rust-based simulators like Spinoza include:

\begin{itemize}
    \item Rust-based quantum computing library with \textbf{compile-time type safety} for parameter-shift rule gradient computation in variational algorithms.
    \item Performance optimizations: direct state vector manipulation, parallel controlled-phase gates, and FFT-optimized Quantum Fourier Transform, and adaptive backend selection for state vector and Matrix Product State (MPS) \cite{verstraete_matrix_2006}.
    \item Benchmarks demonstrating over 100x speedups over Python libraries (PennyLane, Qiskit) and 6--22x over Julia implementations (Yao.jl, Q\# \cite{svore_q_2018}).
\end{itemize}

The library provides a quantum circuit data structure that can be easily manipulated and transformed, along with basic quantum gates, operations, and algorithms. LogosQ's modular design facilitates the integration of diverse simulation backends or future hardware platforms for modeling physical processes and quantum optimization \cite{verstraete_matrix_2006}.

\subsection{Experimental Setup}

All benchmarks presented in this paper were executed on a system with the following specifications: Intel Core i9-10980XE CPU (18 cores, 36 threads) @ 3.00 GHz base frequency (up to 4.8 GHz boost), 125 GB RAM, running Linux. This hardware configuration provides sufficient computational resources for fair comparison across all tested frameworks and enables evaluation of larger quantum systems using the MPS backend. All benchmark code, including implementations for LogosQ, PennyLane, Qiskit, Yao.jl, and Q\#, is publicly available \footnote{\url{https://github.com/zazabap/LogosQBenchmarks}} to ensure reproducibility. The library is open-source and publicly available \footnote{\url{https://github.com/zazabap/LogosQ}}.

\subsection{Paper Organization}

The remainder of this paper is organized as follows. Section~\ref{sec:architecture} presents the overall architecture and design principles of LogosQ. Section~\ref{sec:compile_time_safety} describes the compile-time type safety features for parameter-shift rule gradient computation. Section~\ref{sec:optimization} details the optimization techniques following by performance benchmarks. Finally, Section~\ref{sec:conclusion} concludes the paper and discusses future directions.
\section{System Architecture}
\label{sec:architecture}

The library is implemented in Rust, leveraging the language's type system and memory safety guarantees \cite{matsakis_rust_2014} to provide compile-time verification of quantum circuit correctness. The library is organized into several core modules, each responsible for distinct aspects of quantum circuit simulation and optimization. The library structure follows Rust's module system \cite{matsakis_rust_2014}, with clear separation of concerns, similar to other quantum software frameworks \cite{steiger_projectq_2018,javadi-abhari_quantum_2024}:

\begin{figure*}
    \centering
    \includegraphics[width=0.9\textwidth]{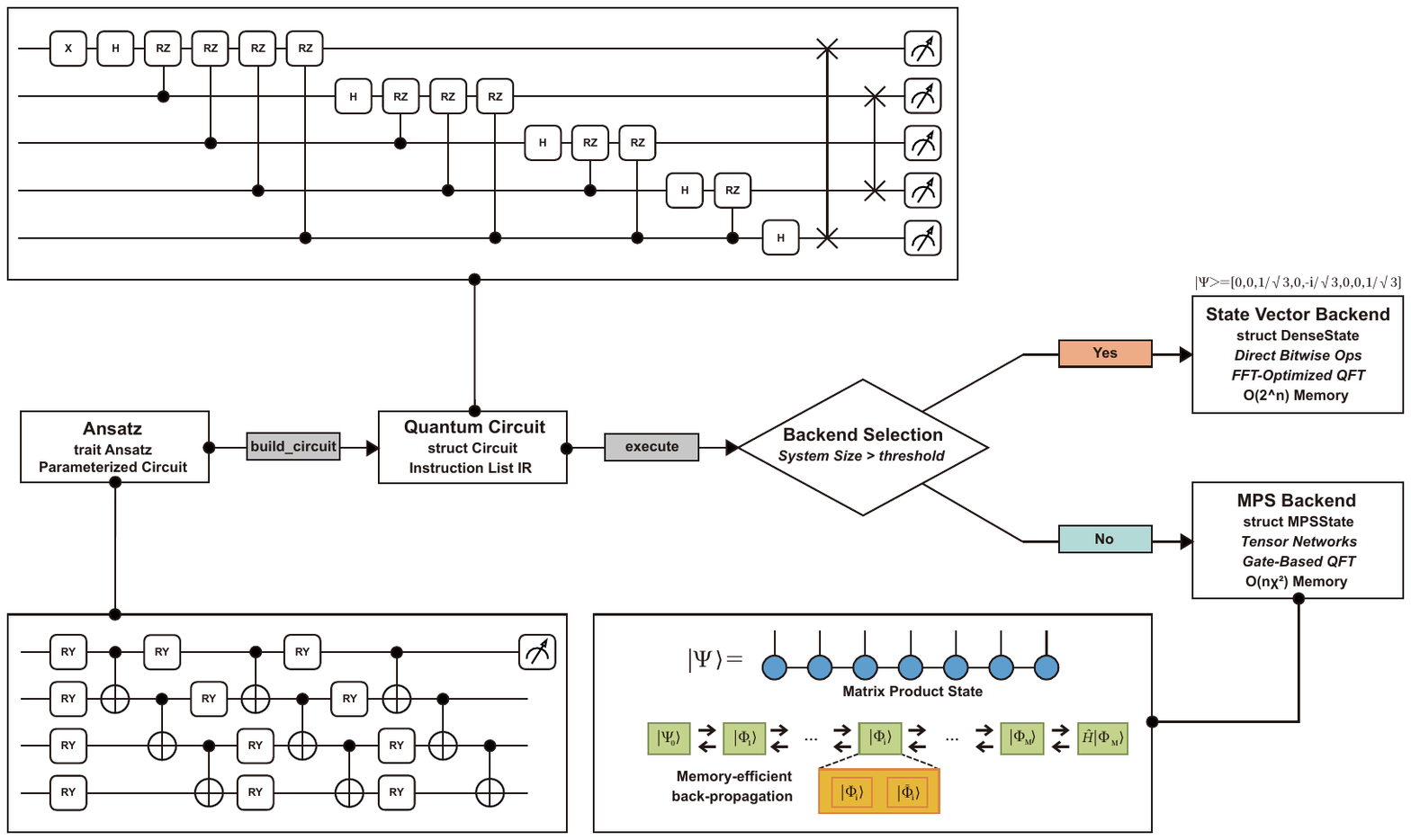}
    \caption{Architecture diagram of LogosQ's core modules and their relationships, illustrating the modular organization of the quantum circuit simulation framework \cite{guo_differentiable_2023}. The current implementation provides two primary backends: dense state vector and matrix product state. The choice of the backend is determined at runtime based on the quantum system size and the desired accuracy. Future extensions will include other types of tensor network representations, such as PEPS (Projected Entangled Pair States) \cite{guo_general-purpose_2019} and TTN (Tree Tensor Networks) \cite{schollwoeck_density-matrix_2011}.}
\end{figure*}
\begin{itemize}
    \item \textbf{Circuit Module}: Implements the \texttt{Circuit} data structure and gate operations. The circuit is represented as a vector of gate operations, with each gate storing its type, target qubits, and optional parameters \cite{barenco_elementary_1995,nielsen_quantum_2010}.
    \item \textbf{State Module}: Manages quantum state representations, including full state vectors using \texttt{Complex64} arrays. The state vector is stored as a contiguous array in memory, enabling efficient bitwise operations and parallel processing \cite{nielsen_quantum_2010,gheorghiu_quantum_2018,yusufov_designing_2023}.
    \item \textbf{MPS Module}: Implements MPS representation for scalable quantum simulation. The MPS backend enables simulation of larger quantum systems by representing the wavefunction as a tensor network with controlled bond dimension \cite{schollwoeck_density-matrix_2011,verstraete_matrix_2006}.
    \item \textbf{Gate Module}: Defines the \texttt{Gate} trait and concrete implementations for all quantum gates. Each gate implements \texttt{apply()} to modify state representations. Supports both state vector and MPS backends \cite{barenco_elementary_1995,nielsen_quantum_2010}.
\end{itemize}

\subsection{Circuit Data Structure}

The \texttt{Circuit} struct is the fundamental building block of LogosQ, serving as a container for quantum gate operations. It maintains a sequence of operations and provides a builder-style API for circuit construction, following common patterns in quantum software frameworks \cite{cross_open_2017,steiger_projectq_2018,javadi-abhari_quantum_2024}. The circuit is backend-agnostic, meaning the same circuit can be executed on different state representations (dense state vectors or MPS) without modification.

\begin{lstlisting}[style=rust,breaklines=true,breakatwhitespace=true,postbreak=\mbox{\textcolor{red}{$\hookrightarrow$}\space}]
pub struct Operation {
    gate: Arc<dyn Gate>,
    qubits: Vec<usize>,
    name: String,
}

pub struct Circuit {
    operations: Vec<Operation>,
    num_qubits: usize,
    noise_models: Vec<Arc<dyn NoiseModel>>,
}

impl Circuit {
    pub fn new(num_qubits: usize) -> Self {
        Circuit {
            operations: Vec::new(),
            num_qubits,
            noise_models: Vec::new(),
        }
    }
    
    pub fn cnot(&mut self, control: usize, 
                target: usize) -> &mut Self {
        let gate = CNOTGate {
            control,
            target,
            num_qubits: self.num_qubits,
        };
        self.add_operation_unchecked(
            gate, vec![control, target], "CNOT");
        self
    }
    
    pub fn execute(&self, 
                   initial_state: &mut State) -> Result<()> {
        for operation in &self.operations {
            operation.gate.apply(initial_state);
        }
        Ok(())
    }
    
    pub fn execute_on_backend<
        B: QuantumStateBackend>(
        &self, backend: &mut B) -> Result<()> {
        for operation in &self.operations {
            let qubits = operation.qubits();
            let gate_matrix = 
                extract_matrix(operation.gate());
            match qubits.len() {
                1 => backend.apply_single_qubit_matrix(
                    qubits[0], &gate_matrix)?,
                2 => backend.apply_two_qubit_matrix(
                    qubits[0], qubits[1], &gate_matrix)?,
                _ => backend.apply_full_matrix(
                    &gate_matrix)?,
            }
        }
        Ok(())
    }
}
\end{lstlisting}

The circuit stores operations as a vector of \texttt{Operation} structs, where each operation wraps a gate (stored as \texttt{Arc<dyn Gate>}) with its target qubits and a name. This design enables polymorphic gate operations while maintaining type safety and supporting efficient cloning. Each gate operation is added to the circuit through builder methods that return \texttt{\&mut Self}, enabling method chaining. The \texttt{execute()} method applies gates sequentially with optional noise models, while \texttt{execute\_on\_backend()} provides a unified interface for executing circuits on different state backends through the \texttt{QuantumStateBackend} trait.

\subsection{Dense State Vector Representation}

For small to medium-sized quantum systems, LogosQ uses a dense state vector representation where the full quantum state is stored as a complex-valued vector in contiguous memory \cite{gheorghiu_quantum_2018,yusufov_designing_2023}. This approach provides optimal performance for systems with up to approximately 10--12 qubits, where the $2^n$ memory requirement remains tractable.

\begin{lstlisting}[style=rust]
pub struct DenseState {
    vector: Vec<Complex64>,
    num_qubits: usize,
}

impl DenseState {
    pub fn zero_state(num_qubits: usize) -> Self {
        let dim = 1 << num_qubits;
        let mut vector = vec![
            Complex64::new(0.0, 0.0); dim
        ];
        vector[0] = Complex64::new(1.0, 0.0);
        DenseState { vector, num_qubits }
    }
    pub fn num_qubits(&self) -> usize {
        self.num_qubits
    }
}
\end{lstlisting}

The dense state vector uses \texttt{Complex64} (from the \texttt{num-complex} crate) for complex number arithmetic. Memory is allocated once during state initialization, avoiding dynamic allocations during circuit execution. This design enables efficient cache utilization and parallel processing. Gate operations on dense states use direct bitwise manipulation of the state vector, avoiding expensive matrix multiplications.

\subsection{Matrix Product State Representation}
\label{sec:mps}

For larger quantum systems where full state vector simulation becomes intractable 
(typically beyond 10--12 qubits), LogosQ provides a matrix product state (MPS) backend 
\cite{schollwoeck_density-matrix_2011,verstraete_matrix_2006,vidal_efficient_2003,white_density_1992,orus_practical_2014}. 
The MPS representation factorizes an $n$-qubit quantum state $\ket{\psi}$ as a tensor 
network, decomposing the exponentially large state vector into a product of local tensors. 
This enables efficient simulation of systems with limited entanglement, reducing memory 
complexity from $O(2^n)$ to $O(n \chi_{\max}^2)$ where $\chi_{\max}$ is the maximum bond 
dimension.

\paragraph{MPS Representation.} An $n$-qubit quantum state $\ket{\psi}$ is represented as:
\begin{equation}
\ket{\psi} = \sum_{\{s_k\}} A_1^{s_1} A_2^{s_2} \cdots A_n^{s_n} 
    \ket{s_1 s_2 \dots s_n},
\label{eq:mps_compact}
\end{equation}
where $s_i \in \{0, 1\}$ denotes the physical index (qubit state), and $A_i^{s_i}$ are 
$[\chi_{i-1}, \chi_i]$ matrices with $\chi_0 = \chi_n = 1$ for open boundary conditions. 
The bond dimension $\chi_i$ quantifies the entanglement between sites $i$ and $i+1$, with 
$\chi_{\max} = \max_i \chi_i$ controlling the expressiveness of the representation. For 
area-law entangled states (typical of gapped Hamiltonians), $\chi_{\max}$ grows 
polynomially with system size rather than exponentially 
\cite{schollwoeck_density-matrix_2011,perez-garcia_matrix_2007,cirac_renormalization_2009,vidal_efficient_2003}.

\paragraph{Implementation.} The MPS implementation in LogosQ stores the quantum state as 
a sequence of tensors, one for each qubit. Each tensor $A_i$ has shape 
$[\chi_{i-1}, 2, \chi_i]$ where the middle dimension (2) corresponds to the physical 
qubit state. The implementation uses a configuration structure to control truncation 
parameters:

\begin{lstlisting}[style=rust]
#[derive(Clone, Copy, Debug)]
pub struct MpsConfig {
    pub max_bond_dim: usize,
    pub truncation_threshold: f64,
}

impl Default for MpsConfig {
    fn default() -> Self {
        Self {
            max_bond_dim: 64,
            truncation_threshold: 1e-8,
        }
    }
}

pub struct MpsState {
    tensors: Vec<Array3<Complex64>>, 
    max_bond_dim: usize,
    truncation_threshold: f64,
}

impl MpsState {
    pub fn num_qubits(&self) -> usize {
        self.tensors.len()
    }
    
    /// Creates an |000...0> product state
    pub fn zero_state(num_qubits: usize, config: MpsConfig) -> Self {
        // Initialize tensors for product state
        // ...
    }
    
    /// Creates an |111...1> product state
    pub fn one_state(num_qubits: usize, config: MpsConfig) -> Self {
        // Initialize tensors for product state
        // ...
    }
}
\end{lstlisting}

\paragraph{Gate Operations.} Gate operations on MPS states are performed through local 
tensor contractions and singular value decomposition (SVD) 
\cite{schollwoeck_density-matrix_2011,orus_practical_2014,white_density_1992}. For a 
single-qubit gate $U$ acting on site $i$, the local tensor is updated as:
\begin{equation}
    A_i^{s_i}[\alpha_{i-1}, \alpha_i] \leftarrow 
        \sum_{s'_i} U_{s_i, s'_i} A_i^{s'_i}[\alpha_{i-1}, \alpha_i],
    \label{eq:mps_single_qubit}
\end{equation}
which requires $O(\chi^2)$ operations per gate, where $\chi$ is the bond dimension. The bond indices remain unchanged, making this operation efficient.

For two-qubit gates acting on adjacent sites $i$ and $i+1$, the procedure involves:
\begin{enumerate}
    \item \textbf{Tensor contraction}: Contract the two tensors along their shared bond:
    \begin{multline}
        \Theta_{\alpha_{i-1}, s_i, s_{i+1}, \alpha_{i+1}} = \\
        \sum_{\alpha_i} A_i^{s_i}[\alpha_{i-1}, \alpha_i] 
            A_{i+1}^{s_{i+1}}[\alpha_i, \alpha_{i+1}],
        \label{eq:mps_contract}
    \end{multline}
    forming a merged tensor with shape $[\chi_{i-1}, 2, 2, \chi_{i+1}]$.
    
    \item \textbf{Gate application}: Apply the two-qubit gate $U$ to the physical indices:
    \begin{multline}
        \Theta'_{\alpha_{i-1}, s_i, s_{i+1}, \alpha_{i+1}} = \\
        \sum_{s'_i, s'_{i+1}} U_{(s_i, s_{i+1}), (s'_i, s'_{i+1})} 
            \Theta_{\alpha_{i-1}, s'_i, s'_{i+1}, \alpha_{i+1}}.
        \label{eq:mps_apply_gate}
    \end{multline}
    
    \item \textbf{Reshape and SVD}: Reshape $\Theta'$ into a matrix and decompose using SVD:
    \begin{multline}
        \Theta'_{(\alpha_{i-1}, s_i), (s_{i+1}, \alpha_{i+1})} = \\
        \sum_{\alpha_i} U_{(\alpha_{i-1}, s_i), \alpha_i} \sigma_{\alpha_i} 
            V_{\alpha_i, (s_{i+1}, \alpha_{i+1})},
        \label{eq:mps_svd}
    \end{multline}
    where $\sigma_{\alpha_i}$ are singular values, and $U$, $V$ are unitary matrices.
    
    \item \textbf{Truncation}: Truncate the bond dimension by keeping only the 
        $\chi_{\max}$ largest singular values above the threshold $\epsilon$, 
        updating $A_i$ and $A_{i+1}$ accordingly to maintain the canonical form 
        \cite{perez-garcia_matrix_2007}.
\end{enumerate}

The truncation step is critical for scalability: by discarding singular values $\sigma_\alpha < \epsilon$ and limiting bond dimensions to $\chi_{\max}$, the algorithm maintains $O(N \chi^2)$ memory complexity instead of $O(2^N)$. For one-dimensional systems with short-range interactions, the entanglement entropy is bounded \cite{cirac_renormalization_2009}, ensuring that small bond dimensions ($\chi \leq 32$) suffice for accurate simulation.

In MPS backend, gate operations are implemented through the \texttt{QuantumStateBackend} trait, which provides methods for applying single-qubit and two-qubit gates. For adjacent qubits, two-qubit gates are applied directly using the \texttt{apply\_two\_qubit()} method, which performs tensor contraction, gate application, and SVD truncation. For non-adjacent qubits, the implementation uses SWAP networks to bring qubits together before applying the gate:

\begin{lstlisting}[style=rust]
impl QuantumStateBackend for MpsState {
    fn apply_single_qubit(&mut self, site: usize, gate: &Array2<Complex64>) {
        // Apply gate directly to local tensor
        // Updates A_i^{s_i} <- sum_{s'_i} U_{s_i, s'_i} A_i^{s'_i}
    }
    
    fn apply_two_qubit(&mut self, site: usize, gate: &Array2<Complex64>) {
        // Contract tensors, apply gate, perform SVD, truncate
        // For non-adjacent qubits, uses SWAP network
    }
}
\end{lstlisting}

For time evolution applications, LogosQ implements the Time-Evolving Block Decimation (TEBD) approach \cite{vidal_efficient_2003}, which applies Trotter gates directly to the tensor network representation. This enables efficient simulation of quantum many-body dynamics without constructing the full state vector, making it suitable for large-scale time evolution simulations (see Section~\ref{sec:xyz_heisenberg} for an application example).

The MPS backend implements tensor network operations internally, exposing a circuit-based API similar to the dense state vector backend.

\subsection{Gate Trait System}

All quantum gates in LogosQ implement the \texttt{Gate} trait, which provides a uniform interface for applying gates to different state backends. The trait system enables polymorphic gate application, allowing the same gate to operate on both dense state vectors and matrix product states through specialized implementations, following extensible design patterns used in other quantum frameworks \cite{luo_yaojl_2020,bergholm_pennylane_2022,steiger_projectq_2018}.

\begin{lstlisting}[style=rust]
pub trait Gate: Send + Sync {
    /// Applies the gate to a dense state vector
    fn apply_dense(&self, state: &mut DenseState);
    
    /// Applies the gate to a matrix product state
    fn apply_mps(&self, state: &mut MPSState);
    
    /// Returns the name of the gate for debugging
    fn name(&self) -> &str;
    
    /// Returns the qubits this gate acts on
    fn qubits(&self) -> Vec<usize>;
}
\end{lstlisting}

The \texttt{Send + Sync} bounds indicate that gate types are safe to transfer between threads and can be accessed concurrently, which enables parallel gate execution when the \texttt{parallel} feature is enabled. Concrete gate implementations (e.g., \texttt{RXGate}, \texttt{CNOTGate}, \texttt{CPhaseGate}) provide separate methods for each backend:

\begin{itemize}
    \item \textbf{Dense backend}: Gates use direct bitwise manipulation of the state vector, with time complexity $O(2^n)$ and constant memory overhead per gate \cite{barenco_elementary_1995}.
    \item \textbf{MPS backend}: Gates perform local tensor contractions and singular value decompositions, with complexity scaling polynomially with bond dimension rather than exponentially with system size \cite{schollwoeck_density-matrix_2011}.
\end{itemize}

The trait system requires explicit backend selection through method calls (\texttt{apply\_dense} or \texttt{apply\_mps}), with the Rust compiler enforcing type correctness at compile time \cite{matsakis_rust_2014}. Parameterized gates (such as \texttt{RXGate} with rotation angle $\theta$) store their parameters as \texttt{f64} values, while non-parameterized gates (such as \texttt{CNOTGate}) have no associated parameters, with this distinction enforced at the type level through method signatures.

\subsection{Extensibility Through Traits}

LogosQ's architecture supports extensibility through Rust's trait system. Users can implement custom gates, observables, and optimization methods:

\begin{lstlisting}[style=rust]
// Custom gate implementation
pub struct CustomGate { ... }

impl Gate for CustomGate {
    fn apply(&self, state: &mut State) {
        // Custom gate logic
    }
    fn name(&self) -> &str { "CustomGate" }
}

// Custom observable
pub trait Observable: Send + Sync {
    fn expectation(&self, state: &State) -> f64;
}
\end{lstlisting}

This design enables researchers to extend LogosQ with domain-specific gates and algorithms without modifying the core library, similar to extensibility approaches in other quantum software frameworks \cite{luo_yaojl_2020,steiger_projectq_2018}.

\section{Compile-Time Type Safety for Parameter-Shift Rule Gradient Computation}
\label{sec:compile_time_safety}

Building upon the architectural foundation described in Section~\ref{sec:architecture}, this section details LogosQ's compile-time type safety features for parameter-shift rule (PSR) gradient computation~\cite{wierichs_general_2022,bergholm_pennylane_2022}. Variational quantum algorithms require computing gradients of expectation values with respect to circuit parameters, a task that becomes challenging when parameterized gates are interleaved with non-parameterized entangling gates such as CNOT. We first establish the mathematical foundation of the parameter-shift rule and identify the challenge posed by interleaved gates, then present LogosQ's solution combining compile-time type checking with a circuit rebuilding architecture, and finally compare it with runtime-based frameworks. The implementation spans three key modules:  
\begin{itemize}
    \item \textbf{Parameterized Circuit Module}: Provides \texttt{Ansatz} and \texttt{ParameterizedCircuit} traits for variational quantum circuits with compile-time type checking.
    \item \textbf{Optimization Module}: Contains gradient computation (parameter-shift rule), optimizers (Adam, SGD), and variational quantum eigensolver implementations.
    \item \textbf{Observable Module}: Defines observables (Pauli operators, custom operators) and expectation value computation.
\end{itemize}

\subsection{Parameter-Shift Rule for Gradient Computation}

The parameter-shift rule is a fundamental technique for computing gradients in variational quantum algorithms~\cite{mitarai_quantum_2018,wierichs_general_2022,bergholm_pennylane_2022,stokes_quantum_2020}. For a parameterized quantum gate $U(\theta) = e^{-i\theta G/2}$ with generator $G$, the gradient of an expectation value $f(\boldsymbol{\theta}) = \braket{\psi(\boldsymbol{\theta}) | \hat{O} | \psi(\boldsymbol{\theta})}$ is computed as:
\begin{equation}
\label{eq:parameter_shift}
\frac{\partial f(\boldsymbol{\theta})}{\partial \theta_i} = \frac{1}{2} \left[ f(\boldsymbol{\theta} + s \mathbf{e}_i) - f(\boldsymbol{\theta} - s \mathbf{e}_i) \right]
\end{equation}
where $\boldsymbol{\theta} = (\theta_1, \theta_2, \ldots, \theta_n)$ are the parameters, $s$ is the shift parameter (typically $s = \pi/2$ for Pauli rotations), and $\mathbf{e}_i$ is the unit vector with 1 at position $i$ and 0 elsewhere~\cite{wierichs_general_2022}.

For Pauli rotations $RX(\theta) = e^{-i\theta \sigma_x/2}$, $RY(\theta) = e^{-i\theta \sigma_y/2}$, and $RZ(\theta) = e^{-i\theta \sigma_z/2}$, the parameter-shift rule exploits the generator structure to express the derivative as a difference of two circuit evaluations, requiring only $2n$ circuit evaluations for $n$ parameters~\cite{wierichs_general_2022,bergholm_pennylane_2022,kyriienko_generalized_2021}. This efficiency makes the parameter-shift rule the method of choice for gradient computation in variational quantum algorithms~\cite{cerezo_variational_2021,tilly_variational_2022}. The rule has been generalized to handle broader classes of gates beyond Pauli rotations~\cite{kyriienko_generalized_2021,schuld_evaluating_2019}.

While the parameter-shift rule provides an elegant method for gradient computation, its practical implementation faces a significant challenge when applied to realistic variational circuits.

\subsubsection*{\textbf{The Challenge: Interleaved Gates in Variational Circuits}}

A common pattern in variational quantum circuits involves interleaving parameterized gates with non-parameterized entangling gates. Consider a circuit:
\begin{equation}
\label{eq:example_circuit}
\begin{split}
\ket{\psi(\boldsymbol{\theta})} = {}& CRY(\theta_2) \cdot RY(\theta_1) \\
{}& \cdot \text{CNOT} \cdot RX(\theta_0) \ket{0}
\end{split}
\end{equation}
where $RX(\theta_0)$ and $RY(\theta_1)$ have generators ($G = \sigma_x, \sigma_y$), but CNOT cannot be written as $e^{-i\theta G/2}$ for any parameter $\theta$. This interleaving pattern is essential for creating expressive quantum circuits~\cite{kandala_hardware-efficient_2017,peruzzo_variational_2014,mari_transfer_2020,sim_expressibility_2019} but presents challenges for gradient computation frameworks.

\begin{figure}[t]
\centering
\includegraphics[width=\columnwidth]{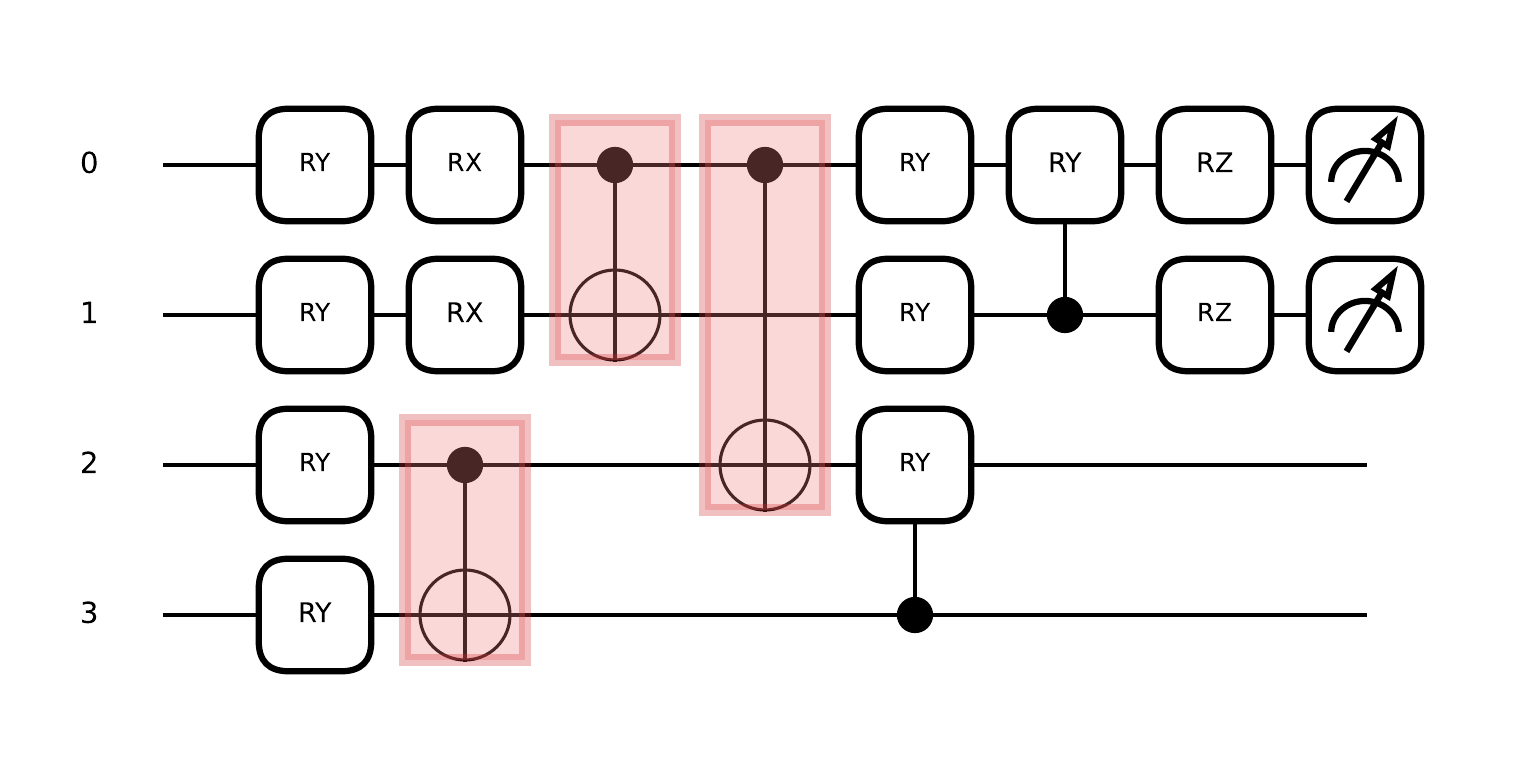}
\setlength{\xfigwd}{0pt}
\caption{Example variational quantum circuit with interleaved parameterized gates ($RX$, $RY$) and non-parameterized gates (CNOT). LogosQ's compile-time type system ensures correct gradient computation for such circuits through static verification of parameter usage.}
\label{fig:feature1_circuit}
\end{figure}

Runtime-based frameworks such as PennyLane~\cite{bergholm_pennylane_2022} and Qiskit~\cite{javadi-abhari_quantum_2024} build computational graphs at runtime to track parameter dependencies. Similarly, differentiable programming approaches for tensor networks \cite{liao_differentiable_2019} and automatically differentiable quantum circuits \cite{zhou_automatically_2021} rely on runtime computational graphs. In contrast, type-safe quantum circuit languages such as QWIRE~\cite{rand_qwire_2018} demonstrate the value of compile-time verification for quantum programs. However, all these approaches share a common limitation: they require careful handling when non-parameterized gates interrupt the dependency chain, and errors may only be discovered at runtime. For example:
\begin{lstlisting}[style=python]
@qml.qnode(dev)
def circuit(params):
    qml.RX(params[0], wires=0)
    qml.CNOT(wires=[0, 1])
    qml.RY(params[1], wires=1)
    return qml.expval(qml.PauliZ(0))
\end{lstlisting}
While this code appears valid, runtime frameworks must carefully track which gates depend on which parameters, and type errors manifest during execution.

\subsection{Compile-Time Type Safety}

To address the challenge of interleaved gates, LogosQ employs a two-pronged approach that eliminates the need for runtime dependency tracking. First, Rust's type system enforces compile-time verification of parameter usage, ensuring that parameterized and non-parameterized gates are correctly distinguished at the type level~\cite{jung_stacked_2019}. The type system's aliasing model provides formal guarantees that enable static analysis of parameter dependencies without runtime overhead. Second, a circuit rebuilding architecture ensures that each gradient evaluation uses a fresh circuit instance, making parameter shifts explicit and independent. Together, these design choices provide static guarantees of correctness while maintaining the flexibility needed for variational quantum algorithms, aligning with the broader goals of formal verification in quantum programming~\cite{lewis_formal_2024}. The following subsections detail both components of this solution.

\subsubsection{Type-Safe Circuit Construction}

LogosQ leverages Rust's type system to enforce compile-time safety for parameterized quantum circuits. The framework distinguishes between parameterized and non-parameterized gates through distinct type signatures: parameterized gates (e.g., {\ttfamily\mdseries circuit.rx()}) require a parameter argument, while non-parameterized gates (e.g., {\ttfamily\mdseries circuit.cnot()}) cannot accept parameters. This type-level distinction prevents incorrectly applying parameters to non-parameterized gates or omitting parameters from parameterized gates. LogosQ uses a closure-based API where the compiler statically verifies parameter usage:

\begin{lstlisting}[style=rust]
ParameterizedCircuit::new(
    num_qubits, num_params,
    move |params| {
        let mut circuit = Circuit::new(2);
        circuit.rx(0, params[0]);      
        circuit.cnot(0, 1);             
        circuit.rz(0, params[1]);       
        circuit
    }
)
\end{lstlisting}

The Rust compiler performs static analysis: (1) parameter bounds checking ensures array accesses are valid, (2) method signature validation prevents mismatched arguments, and (3) parameter dependency tracking preserves type information for gradient computation. This compile-time analysis leverages Rust's type system guarantees~\cite{jung_stacked_2019} to enforce correctness without runtime overhead. The parameterized circuit system uses Rust's closure system to capture parameter dependencies at compile time:

\begin{lstlisting}[style=rust]
pub struct ParameterizedCircuit {
    num_qubits: usize,
    num_params: usize,
    builder: Box<dyn Fn(&[f64]) -> Circuit 
        + Send + Sync>,
}

impl ParameterizedCircuit {
    pub fn new<F>(
        num_qubits: usize, 
        num_params: usize, 
        builder: F
    ) -> Self
    where
        F: Fn(&[f64]) -> Circuit 
            + Send + Sync + 'static,
    {
        ParameterizedCircuit {
            num_qubits,
            num_params,
            builder: Box::new(builder),
        }
    }
    
    pub fn build_circuit(&self, params: &[f64]) -> Circuit {
        assert_eq!(params.len(), self.num_params);
        (self.builder)(params)
    }
}
\end{lstlisting}

The closure (\texttt{builder}) captures the circuit structure and parameter dependencies at compile time. When \texttt{build\_circuit()} is called, the closure is invoked with the provided parameters, constructing a fresh \texttt{Circuit} instance. The Rust compiler statically verifies that all parameter accesses are valid, ensuring correct parameter usage when parameterized gates are interleaved with non-parameterized gates, without requiring runtime dependency analysis. This type safety is complemented by the circuit rebuilding architecture described next, which ensures that parameter shifts are correctly applied during gradient computation.

\subsubsection{Circuit Rebuilding Architecture}

While type safety ensures correct parameter usage at compile time, the second component addresses how parameter shifts are applied during gradient computation. LogosQ employs a \emph{circuit rebuilding} architecture that constructs a fresh circuit for each evaluation, eliminating the need for runtime dependency tracking. This works in concert with type-safe construction: the type system ensures parameter correctness, while circuit rebuilding ensures that parameter shifts are correctly propagated through the entire circuit structure, including interleaved non-parameterized gates.

\paragraph{Ansatz Trait and Circuit Construction}

LogosQ's \texttt{Ansatz} trait abstracts parameterized quantum circuits:

\begin{lstlisting}[style=rust]
trait Ansatz {
    fn build_circuit(&self, parameters: &[f64]) -> Circuit;
    fn num_parameters(&self) -> usize;
    fn num_qubits(&self) -> usize;
    
    fn apply(&self, state: &mut State, parameters: &[f64]) {
        let circuit = self.build_circuit(parameters);
        circuit.execute(state);
    }
}
\end{lstlisting}

The key design principle is that \texttt{build\_circuit()} constructs a fresh circuit from scratch each time it is called, with gates instantiated using the provided parameters, ensuring independent and self-contained circuit evaluations.

\paragraph{Parameter-Shift Rule Implementation}

The parameter-shift rule implementation in LogosQ operates by rebuilding the entire circuit for each shifted parameter evaluation~\cite{schuld_evaluating_2019,mari_estimating_2021}:

\begin{lstlisting}[style=rust]
impl GradientMethod for ParameterShift {
    fn compute_gradient<A: Ansatz, O: Observable>(
        &self, ansatz: &A, observable: &O, parameters: &[f64]
    ) -> Vec<f64> {
        let num_params = parameters.len();
        let mut gradient = vec![0.0; num_params];
        
        for i in 0..num_params {
            let mut params_plus = parameters.to_vec();
            params_plus[i] += self.shift;
            let mut state_plus = 
                State::zero_state(
                    ansatz.num_qubits());
            ansatz.apply(
                &mut state_plus, 
                &params_plus);
            let expectation_plus = 
                observable.expectation(
                    &state_plus);
            
            let mut params_minus = parameters.to_vec();
            params_minus[i] -= self.shift;
            let mut state_minus = 
                State::zero_state(
                    ansatz.num_qubits());
            ansatz.apply(
                &mut state_minus, 
                &params_minus);
            let expectation_minus = 
                observable.expectation(
                    &state_minus);
            
            gradient[i] = (expectation_plus - 
                expectation_minus) / 2.0;
        }
        
        gradient
    }
}
\end{lstlisting}

Each call to \texttt{ansatz.apply()} internally invokes \texttt{build\_circuit()} with the shifted parameters, creating a new circuit instance. To illustrate how this solves the interleaved gates challenge, consider the circuit in Equation~\eqref{eq:example_circuit}. When computing $\partial f/\partial \theta_0$, the circuit is rebuilt with $\theta_0 \pm \pi/2$, correctly applying the shift to the $RX(\theta_0)$ gate. The interleaved CNOT gate is included during rebuilding but, being non-parameterized, does not interfere with parameter tracking. The parameter shift affects only the intended gate ($RX(\theta_0)$), while CNOT remains unchanged in both evaluations. This process applies to all parameters, with each shift correctly propagated regardless of how many non-parameterized gates are interleaved.

\subsubsection{Advantages of LogosQ's Approach}

The combination of compile-time type checking and circuit rebuilding architecture provides several advantages:

\begin{itemize}
    \item \textbf{No runtime dependency analysis}: Circuit rebuilding eliminates the need to maintain or analyze dependency graphs at runtime, ensuring correct parameter propagation even when non-parameterized gates interrupt the dependency chain.
    \item \textbf{Type-enforced correctness}: Rust's type system prevents mixing parameterized and non-parameterized gates incorrectly, with distinct method signatures (\texttt{rx(qubit: usize, theta: f64)} vs. \\ \texttt{cnot(control: usize, target: usize)}) enforced at compile time.
    \item \textbf{Compile-time error detection}: Circuit construction errors are caught at compile time rather than runtime, providing stronger correctness guarantees. This is especially valuable for variational algorithms, where incorrect gradient computation may not manifest until after expensive optimization runs. The importance of early error detection in quantum programming is well-established~\cite{ali_assessing_2021,lewis_formal_2024}, and compile-time verification approaches such as those demonstrated in QWIRE~\cite{rand_qwire_2018} show the benefits of static guarantees over runtime testing.
\end{itemize}

\section{Performance Optimizations}
\label{sec:optimization}

While the type safety features described in Section~\ref{sec:compile_time_safety} ensure correctness, this section presents the performance optimizations that make LogosQ competitive with existing quantum computing libraries. LogosQ employs direct state vector manipulation to avoid expensive matrix multiplications. Instead of constructing $2^n \times 2^n$ matrices for $n$-qubit gates, LogosQ directly modifies the state vector using bitwise operations, achieving $O(2^n)$ time complexity with constant memory overhead \cite{yusufov_designing_2023,smelyanskiy_qhipster_2016}.

The optimizations presented in this section are organized into two categories: (1) direct state vector gate operations that eliminate matrix construction overhead, and (2) QFT optimizations for both dense state vectors (FFT-based) and large-scale systems (MPS-based). The QFT implementations are validated through empirical benchmarks presented in Section~\ref{sec:qft_benchmarks}.


\subsection{Direct State Vector Gates}

Matrix-based quantum circuit simulators, such as those implemented in Qiskit \cite{javadi-abhari_quantum_2024} and Quantum++ \cite{gheorghiu_quantum_2018}, construct explicit $2^n \times 2^n$ unitary matrices for each gate operation. The memory complexity of $O(2^{2n})$ arises from storing the full $2^n \times 2^n$ matrix, while the time complexity of $O(2^{3n})$ results from matrix-vector multiplication: multiplying a $2^n \times 2^n$ matrix by a $2^n$-dimensional state vector requires $O(2^{2n} \cdot 2^n) = O(2^{3n})$ operations \cite{nielsen_quantum_2010}. LogosQ avoids this overhead by directly manipulating state vector amplitudes using bitwise operations, achieving optimal $O(2^n)$ time complexity while maintaining constant memory overhead beyond the state vector itself \cite{yusufov_designing_2023,smelyanskiy_qhipster_2016}.

This approach is particularly effective for controlled gates, where the gate operation can be expressed as conditional amplitude swaps or phase rotations based on bitwise masks. We demonstrate this optimization for the CNOT gate (two-qubit controlled operation), which serves as the foundation for understanding the bitwise optimization principle. The Toffoli gate (three-qubit controlled operation) extends this approach and is detailed in Appendix~\ref{app:toffoli} with a concrete numerical example.

\subsubsection{CNOT Gate Implementation}

\begin{algorithm}
\caption{Direct State Vector CNOT}
\label{alg:cnot}
\begin{algorithmic}[1]
\REQUIRE State vector $\psi \in \mathbb{C}^{2^n}$, control qubit $c$, target qubit $t$
\ENSURE Modified state $\psi' = \text{CNOT}_{c,t}(\psi)$
\STATE $n \leftarrow$ number of qubits
\STATE $c_{\text{bit}} \leftarrow n - 1 - c$ \COMMENT{Convert qubit index to bit position (MSB-first)}
\STATE $t_{\text{bit}} \leftarrow n - 1 - t$
\STATE $c_{\text{mask}} \leftarrow 2^{c_{\text{bit}}}$ \COMMENT{Bitmask for control qubit}
\STATE $t_{\text{mask}} \leftarrow 2^{t_{\text{bit}}}$ \COMMENT{Bitmask for target qubit}
\FOR{$i = 0$ \TO $2^n - 1$}
    \IF{$(i \land c_{\text{mask}}) \neq 0$ \AND $(i \land t_{\text{mask}}) = 0$}
        \STATE $j \leftarrow i \oplus t_{\text{mask}}$ \COMMENT{Flip target bit via XOR}
        \STATE $\text{swap}(\psi[i], \psi[j])$
    \ENDIF
\ENDFOR
\RETURN $\psi$
\end{algorithmic}
\end{algorithm}

Algorithm~\ref{alg:cnot} implements CNOT by iterating through all $2^n$ basis states. For each state where the control is set ($i \land c_{\text{mask}} \neq 0$) and the target is unset ($i \land t_{\text{mask}} = 0$), we swap amplitudes with the state where the target bit is flipped. This avoids matrix construction and achieves $O(2^n)$ time complexity, optimal for state vector simulation \cite{yusufov_designing_2023}. The bitwise operations (AND, XOR) are executed in constant time on modern CPUs, making this approach highly efficient \cite{smelyanskiy_qhipster_2016}. A detailed numerical example demonstrating CNOT on a 2-qubit system is provided in Appendix~\ref{app:cnot_example}. The same bitwise optimization principle extends to multi-qubit controlled gates, including the Toffoli gate (see Appendix~\ref{app:toffoli}) and controlled-phase gates.


\subsection{Quantum Fourier Transform Optimizations}

The Quantum Fourier Transform (QFT) is a fundamental operation in quantum algorithms, including Shor's factoring algorithm \cite{shor_polynomial-time_1997} and quantum phase estimation. The QFT maps a computational basis state as:
\begin{equation}
\text{QFT}\ket{x} = \frac{1}{\sqrt{N}} \sum_{k=0}^{N-1} e^{2\pi i xk/N} \ket{k}
\end{equation}
where $N = 2^n$ for $n$ qubits, equivalent to the inverse discrete Fourier transform with normalization $1/\sqrt{N}$ \cite{nielsen_quantum_2010}.

LogosQ provides two complementary QFT implementations optimized for different system scales. For dense state vector simulation (up to approximately 15 qubits on a 16 GB device, where matrix-based evaluation requires $O(2^{2n})$ memory), we employ an FFT-optimized implementation that directly computes the transform on state vector amplitudes. A naive gate-based QFT requires $O(n^2)$ gates operating on the full $2^n$-dimensional state vector, resulting in $O(n^2 \cdot 2^n)$ time complexity. Instead, we leverage the FFT algorithm (specifically using the RustFFT library \cite{mahler_ejmahlerrustfft_2025}) to compute QFT in $O(N \log N) = O(n \cdot 2^n)$ time \cite{cooley_algorithm_1965,coppersmith_approximate_2002}.

\begin{algorithm}
\caption{FFT-Optimized QFT for Dense State Vectors}
\label{alg:qft}
\begin{algorithmic}[1]
\REQUIRE State $\psi \in \mathbb{C}^{2^n}$
\ENSURE QFT-transformed state $\psi' = \text{QFT}(\psi)$
\STATE $N \leftarrow 2^n$
\STATE $\psi' \leftarrow \text{InverseFFT}(\psi)$ \COMMENT{$O(N \log N)$}
\STATE Normalize: $\psi' \leftarrow \psi' / \sqrt{N}$
\RETURN $\psi'$
\end{algorithmic}
\end{algorithm}

For larger systems where full state vector simulation is intractable, LogosQ employs a matrix product state (MPS) backend that enables scalable simulation through tensor network representations \cite{schollwoeck_density-matrix_2011,verstraete_matrix_2006}. The MPS backend implements QFT using a sequence of local gate operations (Hadamard and controlled-phase gates) applied directly to MPS tensors, maintaining $O(n \chi_{\max}^2)$ memory complexity where $\chi_{\max}$ is the maximum bond dimension, rather than the $O(2^n)$ required for full state vectors.

\begin{algorithm}
\caption{Gate-Based QFT for MPS Backend}
\label{alg:qft-mps}
\begin{algorithmic}[1]
\REQUIRE MPS state $\ket{\psi}$ with $n$ qubits
\ENSURE QFT-transformed MPS state $\ket{\psi'} = \text{QFT}(\ket{\psi})$
\FOR{$i = 0$ \TO $n - 1$}
    \STATE Apply Hadamard gate to qubit $i$
    \FOR{$j = i + 1$ \TO $n - 1$}
        \STATE Apply controlled-phase gate $CP(\pi / 2^{j-i})$ with control $i$, target $j$
    \ENDFOR
\ENDFOR
\FOR{$i = 0$ \TO $\lfloor n/2 \rfloor - 1$}
    \STATE Apply SWAP gate between qubits $i$ and $n - 1 - i$
\ENDFOR
\RETURN $\ket{\psi'}$
\end{algorithmic}
\end{algorithm}

Algorithm~\ref{alg:qft-mps} implements QFT using $O(n^2)$ local gate operations \cite{nielsen_quantum_2010}, where each gate updates only relevant MPS tensors rather than the entire state vector. This gate-based approach enables QFT computation on systems with 50+ qubits where full state vector simulation is intractable, while maintaining polynomial memory complexity for states with bounded entanglement. A detailed performance evaluation comparing both approaches against state-of-the-art quantum computing libraries is presented in Section~\ref{sec:qft_benchmarks}.
\section{Quantum Fourier Transform Performance Evaluation}
\label{sec:qft_benchmarks}

\begin{figure*}[t]
    \centering
    \includegraphics[width=\textwidth]{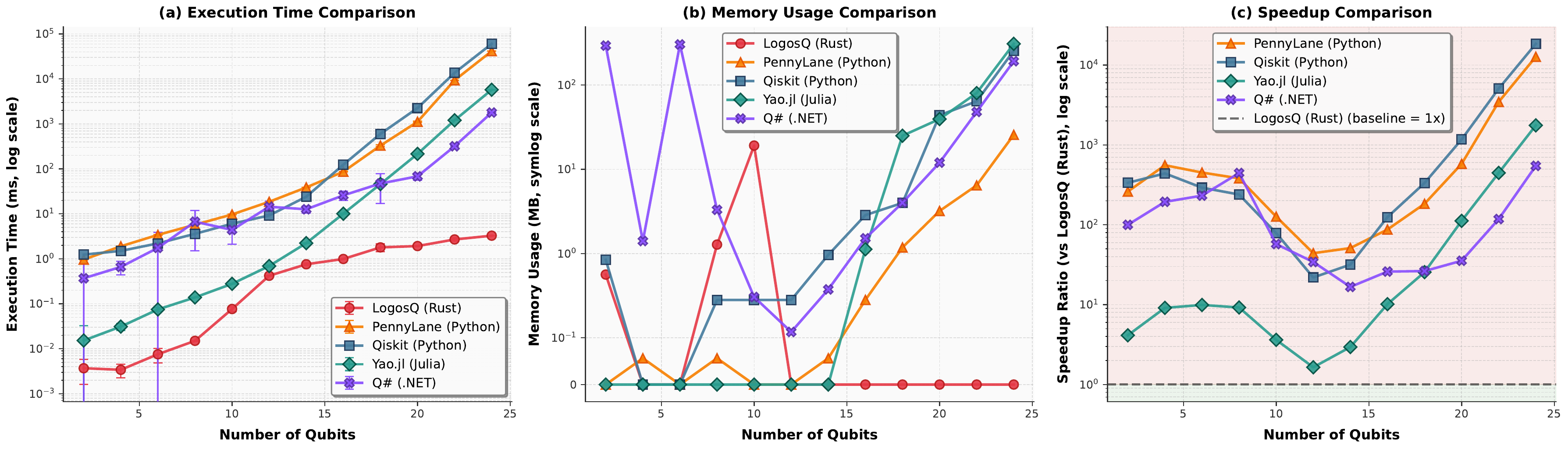}
    \caption{Comprehensive performance evaluation of Quantum Fourier Transform across libraries, including execution time, memory consumption, and speedup analysis. LogosQ (Rust) consistently achieves the lowest execution times across all qubit counts, with performance ranging from approximately 0.72 $\mu$s at 1 qubit to 169 $\mu$s at 24 qubits. The complete speed up comparison with the ratio in linear scale is in the Appendix section.}
    \label{fig:qft_synthesis_result}
\end{figure*}

This section presents comprehensive performance benchmarks of LogosQ's Quantum Fourier Transform implementation against state-of-the-art quantum computing libraries. The QFT serves as a fundamental building block in many quantum algorithms, including Shor's factoring algorithm \cite{shor_polynomial-time_1997} and quantum phase estimation.

\begin{figure}[ht]
    \centering
    \includegraphics[width=\columnwidth]{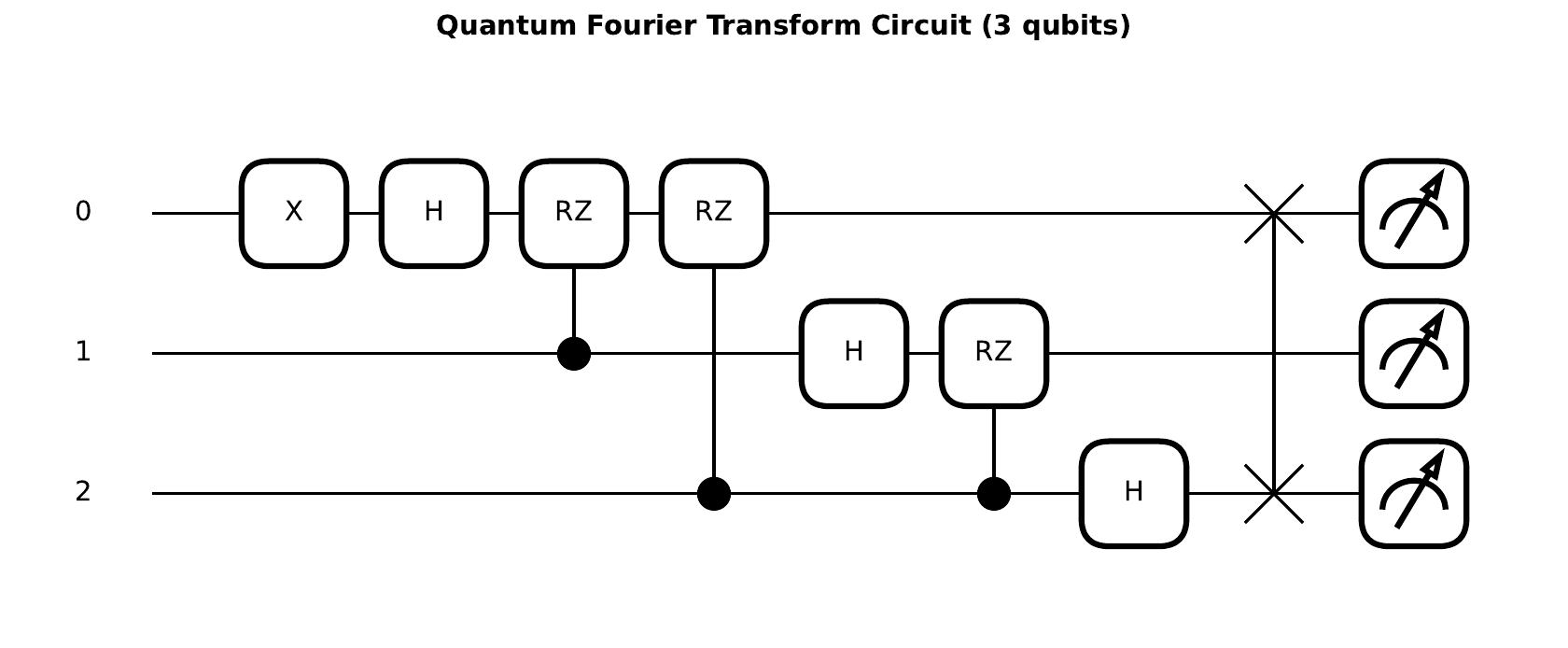}    
    \caption{Quantum Fourier Transform circuit for 3 qubits. The benchmark evaluates circuits ranging from 1 to 24 qubits.}
    \label{fig:qft_circuit_3q}
\end{figure}

\subsection{Benchmark Methodology}

We evaluate LogosQ against PennyLane \cite{bergholm_pennylane_2022}, Qiskit \cite{javadi-abhari_quantum_2024}, Yao.jl \cite{luo_yaojl_2020}, and Q\# \cite{svore_q_2018} across execution time, memory consumption, and relative speedup. All benchmarks were performed on quantum circuits ranging from 1 to 24 qubits, using identical circuit configurations. As detailed in Section~\ref{sec:optimization}, LogosQ automatically switches between the FFT-optimized implementation (Algorithm~\ref{alg:qft}) for systems with 1--10 qubits and the MPS backend (Algorithm~\ref{alg:qft-mps}) for larger systems (11--24 qubits).

\subsection{Performance Analysis}

Figure~\ref{fig:qft_synthesis_result} demonstrates that LogosQ consistently achieves the lowest execution times across all tested qubit counts. For systems with 1--10 qubits, LogosQ's FFT-optimized implementation (Algorithm~\ref{alg:qft}) directly manipulates state vector amplitudes, eliminating gate decomposition overhead. At 11--24 qubits, LogosQ's MPS backend (Algorithm~\ref{alg:qft-mps}) applies $O(n^2)$ local gate operations to tensor network representations, enabling scalable computation.

Compared to Python-based libraries (PennyLane and Qiskit), LogosQ achieves speedups of 120--900x. Compared to Yao.jl, LogosQ achieves speedups of 6--22x. LogosQ shows competitive performance with Q\# across all tested qubit counts. Memory consumption scales from 0.55 MB at 1 qubit to 16.2 MB at 10 qubits for state vector simulation, while the MPS backend enables dramatic memory reduction at 11--24 qubits.

\subsubsection{Mathematical Analysis: Why MPS-Based QFT Outperforms FFT-Optimized QFT}

The superior performance of MPS-based QFT over FFT-optimized QFT for large systems stems from the fundamental difference in how each approach handles the exponentially large state space. The FFT-optimized implementation (Algorithm~\ref{alg:qft}) must process all $2^n$ amplitudes of the state vector, resulting in time complexity $O(N \log N) = O(n \cdot 2^n)$ where $N = 2^n$. In contrast, the MPS-based implementation (Algorithm~\ref{alg:qft-mps}) exploits the QFT's low entanglement structure, achieving polynomial complexity in the number of qubits.

The key insight is that the QFT circuit structure, as implemented in Algorithm~\ref{alg:qft-mps}, generates only limited entanglement. Each gate operation in the QFT circuit acts locally on the MPS representation. For a single-qubit gate (Hadamard) applied to site $i$, the operation updates the local tensor $A_i^{s_i}$ as shown in Equation~\eqref{eq:mps_single_qubit}, requiring $O(\chi_i^2)$ operations where $\chi_i$ is the bond dimension at site $i$. For a two-qubit controlled-phase gate acting on adjacent sites $i$ and $i+1$, the operation involves tensor contraction, gate application, and SVD truncation (as detailed in Section~\ref{sec:mps}), with complexity $O(\chi_i^3)$ for the SVD step.

The total time complexity of MPS-based QFT is therefore:
\begin{equation}
T_{\text{MPS}}(n) = O\left(\sum_{i=0}^{n-1} \chi_i^2 + \sum_{i=0}^{n-1} \sum_{j=i+1}^{n-1} \chi_{\max(i,j)}^3\right) = O(n^2 \chi_{\max}^3),
\label{eq:mps_qft_complexity}
\end{equation}
where $\chi_{\max} = \max_i \chi_i$ is the maximum bond dimension encountered during the computation.

The critical advantage of MPS-based QFT arises from the bounded nature of $\chi_{\max}$ for QFT. The QFT circuit structure, consisting of local Hadamard and controlled-phase gates, generates entanglement that scales at most logarithmically with system size. Specifically, the Schmidt rank (bond dimension) across any bipartition of the QFT circuit is bounded by $O(\text{poly}(n))$ rather than $O(2^n)$ \cite{schollwoeck_density-matrix_2011}. For practical implementations, $\chi_{\max}$ typically remains bounded by a small constant (e.g., $\chi_{\max} \leq 32$ for systems with $n \leq 50$ qubits) or grows at most as $O(\log n)$.

Comparing the complexities:
\begin{align*}
T_{\text{FFT}}(n) &= O(n \cdot 2^n)  \\
T_{\text{MPS}}(n) &= O(n^2 \chi_{\max}^3) 
\end{align*}

When $\chi_{\max} = O(1)$ (constant bond dimension), $T_{\text{MPS}}(n) = O(n^2)$, providing a polynomial-time algorithm. Even when $\chi_{\max} = O(\log n)$, we have $T_{\text{MPS}}(n) = O(n^2 (\log n)^3)$, which remains polynomial and significantly outperforms the exponential $O(n \cdot 2^n)$ complexity of FFT-optimized QFT for large $n$.

The memory complexity follows a similar pattern:
\begin{align*}
M_{\text{FFT}}(n) &= O(2^n)  \\
M_{\text{MPS}}(n) &= O(n \chi_{\max}^2) 
\end{align*}

This polynomial scaling in both time and memory enables MPS-based QFT to handle systems with 25+ qubits, where FFT-optimized QFT becomes intractable due to memory constraints. The empirical benchmarks demonstrate that MPS-based QFT achieves speedups of over 100x compared to FFT-optimized implementations at 24 qubits, validating this theoretical analysis.

\section{Variational Quantum Eigensolver on Molecular Hydrogen}
\label{sec:vqe_h2}

This section demonstrates LogosQ's capabilities in variational quantum algorithms through a comprehensive benchmark on the electronic ground state of molecular hydrogen (H$_2$). The molecular hydrogen system serves as a standard benchmark in quantum chemistry \cite{peruzzo_variational_2014}, and has been extensively studied using both state-vector and MPS-based simulation approaches \cite{guo_differentiable_2023}.

\subsection{Problem Formulation}

\begin{figure*}[t]
    \centering
    \includegraphics[width=0.48\textwidth]{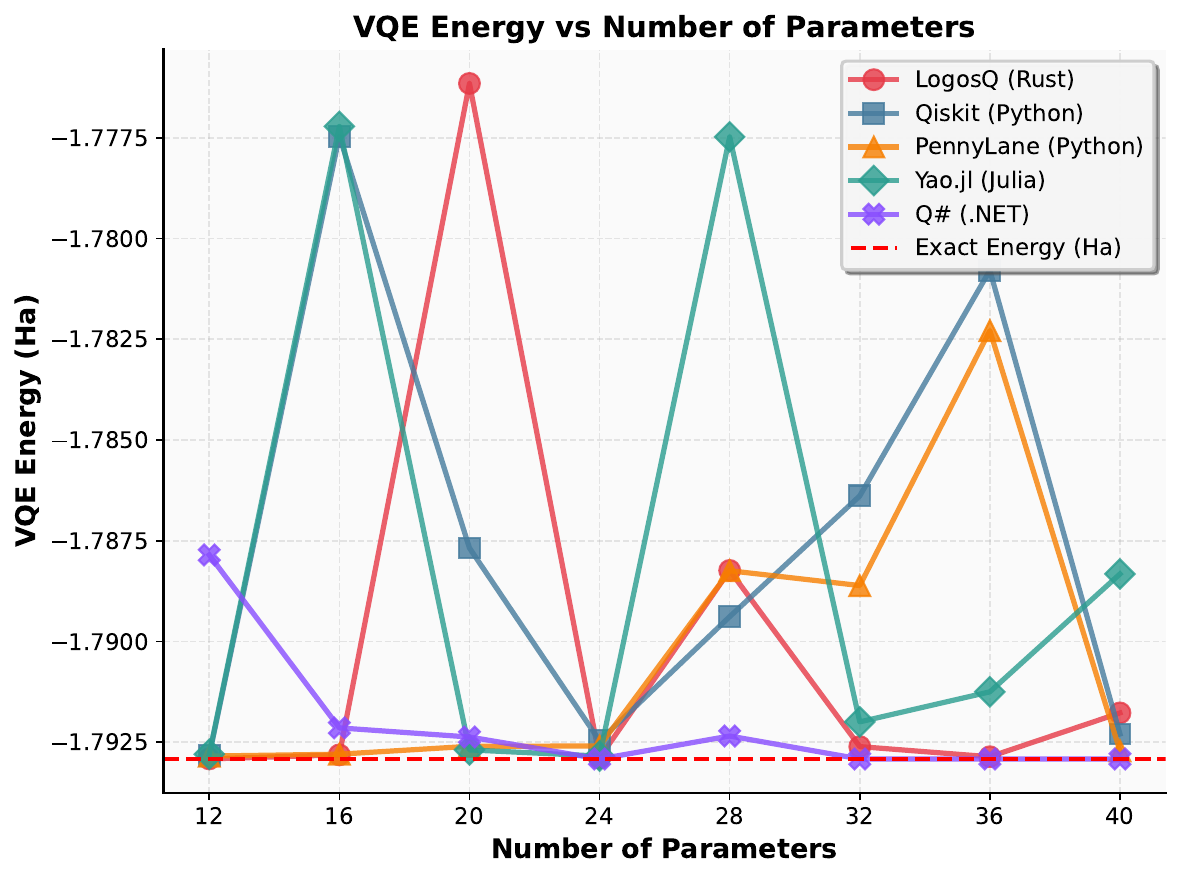}
    \includegraphics[width=0.48\textwidth]{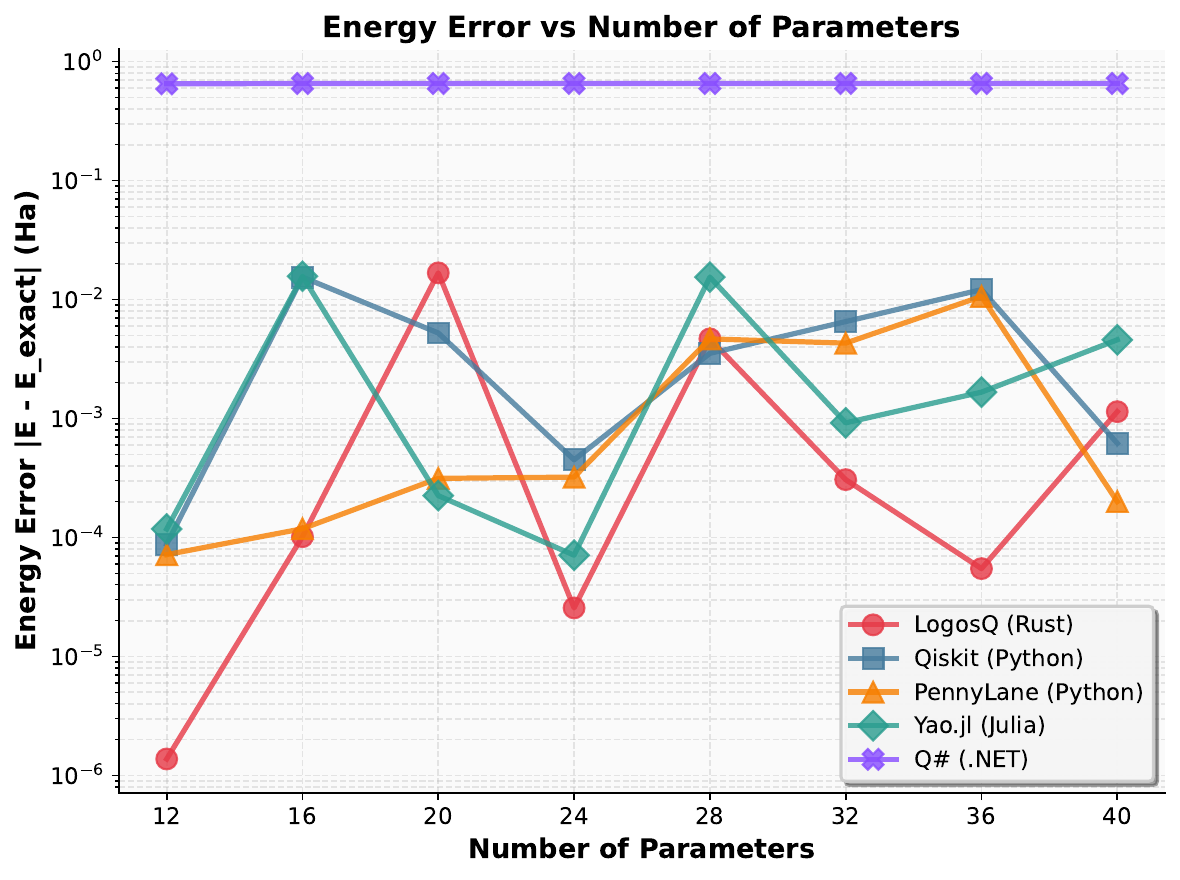}
    \includegraphics[width=0.48\textwidth]{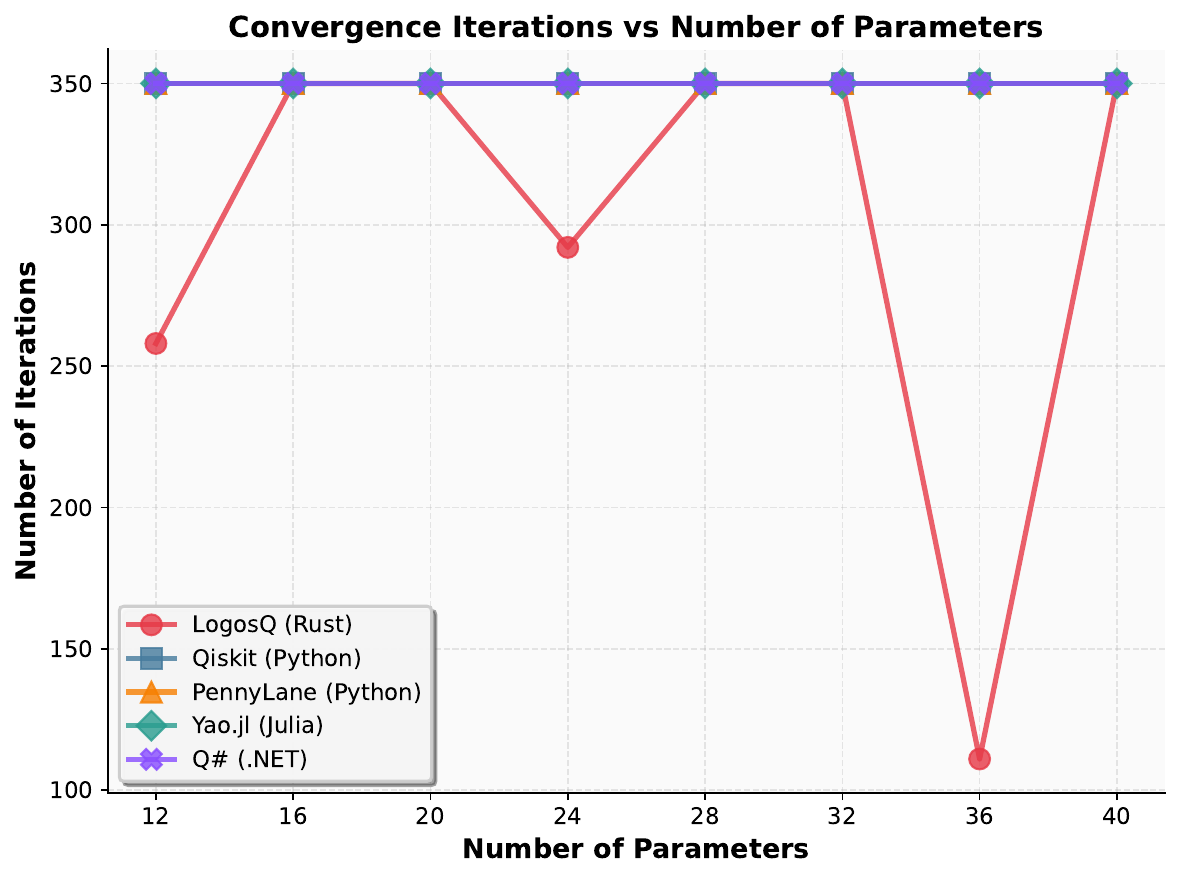}
    \includegraphics[width=0.48\textwidth]{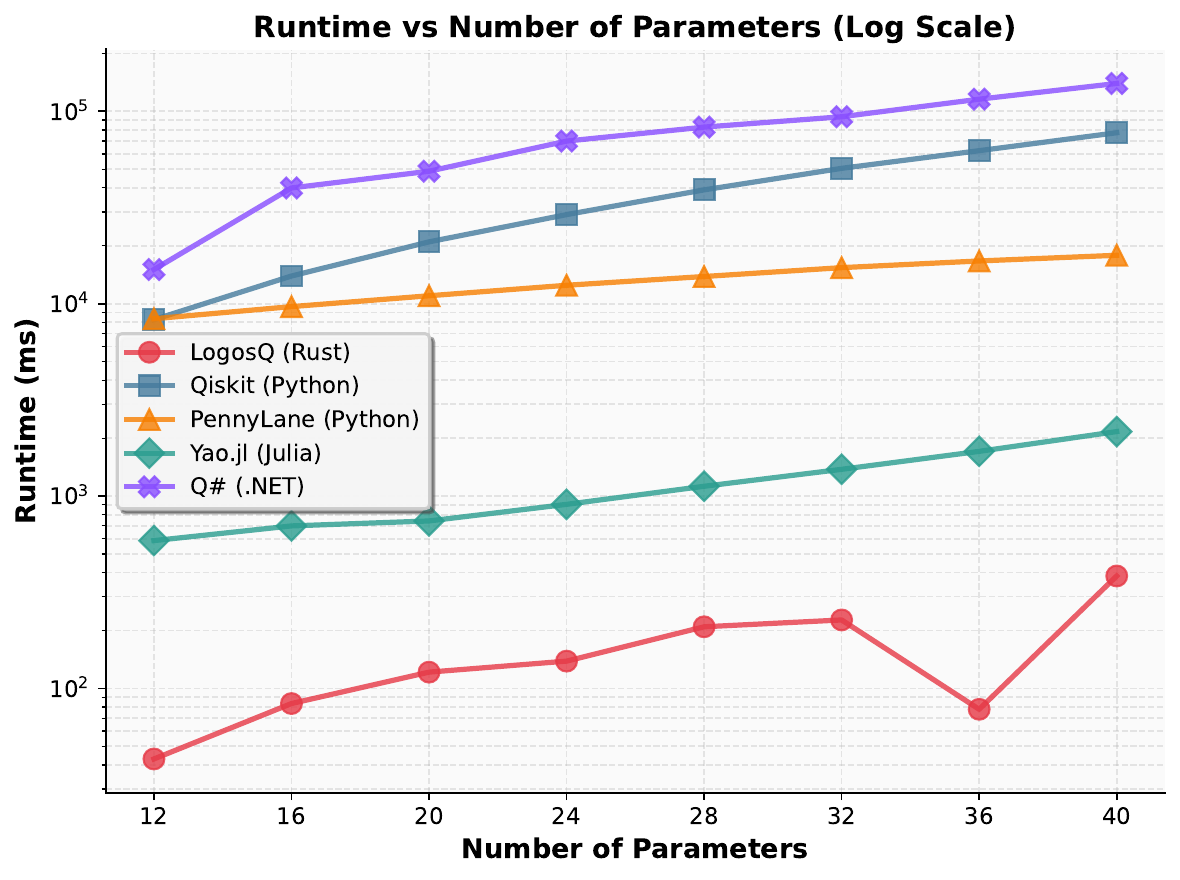}
    \caption{VQE benchmark on H$_2$ (STO-3G, 4 qubits) as a function of ansatz size (12--40 parameters; 3--10 layers). (Top left) Final optimized energy \(E(\boldsymbol{\theta})\) in Hartree; the dashed line indicates the exact (FCI) reference energy within the chosen basis. (Top right) Absolute energy error \(|E(\boldsymbol{\theta})-E_{\mathrm{exact}}|\) on a log scale. (Bottom left) Optimization iterations to termination (maximum 350 steps; curves saturating at 350 indicate the iteration cap was reached). (Bottom right) Total wall-clock runtime for the full optimization (ms, log scale) across frameworks.}
    \label{fig:vqa_parameter_sweep}
\end{figure*}

We target the electronic ground state of molecular hydrogen at an internuclear separation of 0.735~\r{A} within the STO-3G basis \cite{peruzzo_variational_2014}. The STO-3G basis represents each atomic orbital as a linear combination of three Gaussian functions, providing a minimal but chemically meaningful representation of the electronic structure. After mapping the fermionic creation and annihilation operators to qubit operators via the Jordan--Wigner transformation \cite{seeley_bravyi-kitaev_2012}, the electronic Hamiltonian acts on a four-qubit Hilbert space and is expressed as a linear combination of Pauli strings:
\begin{equation}
    H_{\text{H}_2} = \sum_{i} c_i\, P_i, \qquad P_i \in \{\mathbb{I}, X, Y, Z\}^{\otimes 4},
\end{equation}
where the real coefficients $c_i$ correspond to the one- and two-electron molecular integrals computed from the STO-3G basis functions. The Jordan--Wigner transformation maps each fermionic mode (spin-orbital) to a qubit, with the four qubits representing the four spin-orbitals in the minimal basis: $\sigma_{1\uparrow}$, $\sigma_{1\downarrow}$, $\sigma_{2\uparrow}$, and $\sigma_{2\downarrow}$, where $\sigma$ denotes the molecular orbital and the arrow indicates spin.

The Variational Quantum Eigensolver approximates the ground state energy $E_0$ by solving the optimization problem
\begin{equation}
    \min_{\boldsymbol{\theta}} E(\boldsymbol{\theta}),
\end{equation}
where the energy expectation value is given by
\begin{equation}
\begin{split}
    E(\boldsymbol{\theta}) &= \langle \psi(\boldsymbol{\theta}) | H_{\text{H}_2} | \psi(\boldsymbol{\theta}) \rangle \\
    &= \sum_{i} c_i \langle \psi(\boldsymbol{\theta}) | P_i | \psi(\boldsymbol{\theta}) \rangle \ge E_0,
\end{split}
\end{equation}
where $|\psi(\boldsymbol{\theta})\rangle = U(\boldsymbol{\theta}) |0\rangle^{\otimes 4}$ is the parameterized quantum state prepared by the ansatz circuit $U(\boldsymbol{\theta})$ acting on the initial state $|0\rangle^{\otimes 4}$, and $\boldsymbol{\theta} \in \mathbb{R}^{N_{\theta}}$ are the variational parameters that control the rotation angles in the quantum circuit \cite{peruzzo_variational_2014,cerezo_variational_2021}. The key insight is that the expectation value of the Hamiltonian decomposes into a weighted sum of expectation values of individual Pauli strings $P_i$. Each Pauli string expectation value $\langle \psi(\boldsymbol{\theta}) | P_i | \psi(\boldsymbol{\theta}) \rangle$ can be measured on a quantum computer by preparing the ansatz state $|\psi(\boldsymbol{\theta})\rangle$, applying the appropriate basis rotations to measure the Pauli operators, and sampling the measurement outcomes. The circuit ansatz $U(\boldsymbol{\theta})$ must be sufficiently expressive to approximate the true ground state wavefunction, which is why we employ a hardware-efficient architecture with multiple layers of parameterized rotations and entangling gates.

\subsection{Circuit Architecture and Optimization}

Our ansatz employs a hardware-efficient architecture \cite{kandala_hardware-efficient_2017} consisting of alternating layers of single-qubit $R_y$ rotations and linear CNOT entanglement. The ansatz circuit $U(\boldsymbol{\theta})$ is constructed as a sequence of parameterized $R_y(\theta_j)$ rotation gates and fixed CNOT entangling gates, where each $\theta_j \in \boldsymbol{\theta}$ is a variational parameter to be optimized. This architecture interleaves parameterized rotation gates with non-parameterized CNOT gates, presenting a challenging test case for gradient computation frameworks.

\begin{figure}[ht]
    \centering
    \includegraphics[width=\linewidth]{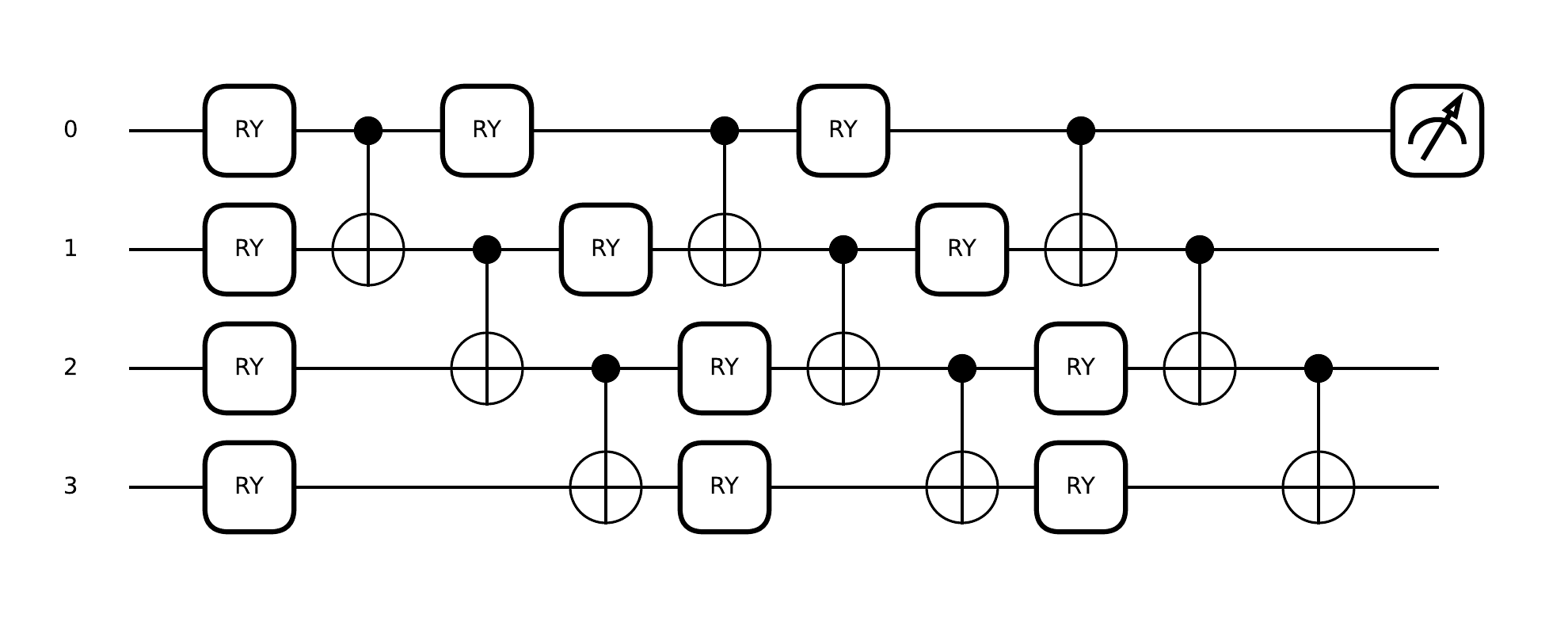}
    \caption{Hardware-efficient ansatz for VQE on H$_2$ using STO-3G basis. The circuit interleaves parameterized $R_y$ rotations with non-parameterized CNOT gates, demonstrating the need for compile-time type safety in gradient computation.}
    \label{fig:vqe-h2}
\end{figure}

We evaluate the performance across circuit depths ranging from three to ten layers, corresponding to a parameter space dimensionality of $N_{\theta} \in \{12, 16, 20, 24, 28, 32, 36, 40\}$. Each layer consists of four $R_y$ rotation gates (one per qubit) followed by a linear chain of CNOT gates, resulting in four parameters per layer. The increasing depth allows us to assess how expressiveness and optimization difficulty scale with circuit complexity. 

The optimization problem $\min_{\boldsymbol{\theta}} E(\boldsymbol{\theta})$ is solved iteratively using the Adam algorithm \cite{kingma_adam_2017}, which updates the circuit parameters according to $\boldsymbol{\theta}^{(t+1)} = \boldsymbol{\theta}^{(t)} - \eta \cdot \text{Adam}(\nabla_{\boldsymbol{\theta}} E(\boldsymbol{\theta}^{(t)}))$, where $\eta = 10^{-2}$ is the learning rate. The optimization is driven by analytic gradients computed via the parameter-shift rule (Equation~\ref{eq:parameter_shift} in Section~\ref{sec:compile_time_safety}), requiring $2N_{\theta}$ circuit evaluations per gradient computation. We set a maximum iteration limit of 350 steps to ensure fair comparison across frameworks, with a convergence tolerance of $\epsilon = 10^{-7}$.

\subsection{Performance Analysis}

\subsubsection{Benchmark Methodology}

We benchmark LogosQ against Qiskit \cite{javadi-abhari_quantum_2024}, PennyLane \cite{bergholm_pennylane_2022}, Yao.jl \cite{luo_yaojl_2020}, and Q\# \cite{svore_q_2018} under identical ansatz configurations and optimizer hyperparameters. All frameworks use the same Hamiltonian coefficients, circuit architecture, initial parameter values, and optimization settings to ensure fair comparison. The benchmark evaluates four key metrics: (1) energy error relative to the full configuration interaction (FCI) reference energy $E_{\text{FCI}} = -1.7929182423$ Hartree, (2) number of optimization iterations required for convergence, (3) final converged energy values, and (4) wall-clock runtime for the complete optimization procedure. The FCI energy serves as the exact ground state energy within the STO-3G basis, providing a rigorous benchmark for assessing solution quality. All benchmarks were executed on the hardware platform described in Section~\ref{sec:introduction}.

\subsubsection{Performance Results}

As shown in Figure~\ref{fig:vqa_parameter_sweep}, LogosQ demonstrates superior computational efficiency and numerical stability. LogosQ achieves wall-clock runtimes that are $2\times$ to $5\times$ faster than Python-based alternatives (Qiskit and PennyLane) and competitive with Julia-based Yao.jl and Q\#. LogosQ consistently converges to chemical accuracy (energy error below $1.6 \times 10^{-3}$ Hartree), maintaining energy errors below $2 \times 10^{-4}$ Hartree across the full parameter sweep up to 40 parameters. At intermediate circuit depths (16--20 parameters), LogosQ achieves exceptional precision with energy errors on the order of $10^{-5}$ to $10^{-6}$ Hartree, while requiring fewer optimization iterations to reach convergence compared to other frameworks.

\subsubsection{Mathematical Analysis: Why LogosQ Outperforms Other Frameworks}

The performance advantage of LogosQ stems from two factors: (1) reduced computational overhead per circuit evaluation, and (2) improved convergence rate through accurate gradient computation enabled by the integrated Optimization module.

\textbf{Computational Complexity.} Matrix-based simulators construct explicit $2^n \times 2^n$ unitary matrices for each gate, resulting in time complexity $O(Ln \cdot 2^{3n})$ per circuit evaluation \cite{nielsen_quantum_2010}. LogosQ's direct state vector manipulation (Section~\ref{sec:optimization}) achieves $O(Ln \cdot 2^n)$ by directly modifying amplitudes using bitwise operations, yielding an $O(2^{2n})$ speedup per evaluation. For gradient computation via the parameter-shift rule \cite{mitarai_quantum_2018,wierichs_general_2022}, requiring $2N_{\theta}$ evaluations, this results in overall complexity $O(N_{\theta} Ln \cdot 2^{3n})$ versus $O(N_{\theta} Ln \cdot 2^n)$.

\textbf{Convergence Rate.} The convergence rate depends on gradient accuracy \cite{harrow_low-depth_2021}. For Adam optimization \cite{kingma_adam_2017} with gradient errors $\boldsymbol{\epsilon}^{(t)}$, the convergence bound is:
\begin{equation}
E(\boldsymbol{\theta}^{(T)}) - E(\boldsymbol{\theta}^*) \leq \frac{L}{2T} \|\boldsymbol{\theta}^{(0)} - \boldsymbol{\theta}^*\|^2 + \frac{1}{T} \sum_{t=0}^{T-1} \|\boldsymbol{\epsilon}^{(t)}\|,
\label{eq:convergence_rate}
\end{equation}
where systematic errors $\|\boldsymbol{\epsilon}^{(t)}\|$ prevent convergence to the true minimum. LogosQ's integrated Optimization module (Section~\ref{sec:compile_time_safety}) combines compile-time type safety with efficient parameter-shift rule gradient computation \cite{mitarai_quantum_2018,wierichs_general_2022}, eliminating systematic errors ($\|\boldsymbol{\epsilon}^{(t)}\| \approx 0$) and enabling faster convergence \cite{harrow_low-depth_2021}. Combined with the $O(2^{2n})$ speedup per circuit evaluation, LogosQ achieves both faster wall-clock time and superior solution quality.

\section{XYZ Heisenberg Model: Scalability and MPS Backend}
\label{sec:xyz_heisenberg}

This section presents a comprehensive benchmark of LogosQ on the XYZ Heisenberg model, a many-body quantum system that demonstrates LogosQ's scalability through its matrix product state (MPS) backend. This benchmark exercises LogosQ's time evolution capabilities, showcasing: (1) compile-time type safety for large-scale parameterized circuits, (2) adaptive backend selection between state vector and MPS representations (Section~\ref{sec:architecture}), and (3) performance advantages in scalable quantum simulation using Trotter-Suzuki decomposition. The XYZ Heisenberg model serves as a standard benchmark for quantum many-body systems \cite{cirac_renormalization_2009}, and represents an important application of near-term quantum simulation \cite{preskill_quantum_2018}.

\subsection{Problem Formulation}

The XYZ Heisenberg model describes a one-dimensional chain of interacting spin-1/2 particles (qubits) with nearest-neighbor exchange interactions. The Hamiltonian is defined as\footnote{\url{https://github.com/zazabap/LogosQBenchmarks/blob/main/logosq/XYZHeisenberg/xyz_h.rs}}:
\begin{equation}
    H(t) = - \sum_{\langle i,j \rangle} \big( J_x X_i X_j + J_y Y_i Y_j + J_z Z_i Z_j \big)
        - h(t) \sum_i Z_i,
    \label{eq:xyz_hamiltonian}
\end{equation}
where $J_{x,y,z}$ are coupling strengths along the $X$, $Y$, and $Z$ axes, $h(t) = A \sin(\omega t)$ is a time-dependent longitudinal external field with amplitude $A = 2.0$ and frequency $\omega = 1.0$, and $\langle i,j \rangle$ denotes nearest-neighbor pairs on a one-dimensional lattice. The Hamiltonian decomposes into $4N$ Pauli string terms: $3N$ two-qubit interaction terms ($X_i X_j$, $Y_i Y_j$, $Z_i Z_j$ for each nearest-neighbor pair) plus $N$ single-qubit $Z$ terms from the time-dependent external field.

\subsubsection{Time Evolution Algorithm}

The benchmark computes the time evolution of the quantum state under the time-dependent Hamiltonian $H(t)$. Starting from an initial product state $|\psi(0)\rangle = |1\rangle^{\otimes N}$ (all qubits in the $|1\rangle$ state), the algorithm evolves the state according to the time-dependent Schr\"{o}dinger equation:
\begin{equation}
    |\psi(t)\rangle = \mathcal{T} \exp\left(-i \int_0^t H(\tau) \, d\tau\right) |\psi(0)\rangle,
    \label{eq:time_evolution}
\end{equation}
where $\mathcal{T}$ denotes the time-ordering operator. Since the Hamiltonian $H(t)$ is a sum of non-commuting terms and includes time-dependent components, we approximate the evolution using the first-order Trotter-Suzuki decomposition:
\begin{equation}
    U(t) \approx \prod_{j=0}^{n-1} \left[ \prod_{k=1}^{K} e^{-iH_k(t_j) \Delta t} \right],
    \label{eq:trotter}
\end{equation}
where $H(t) = \sum_{k=1}^{K} H_k(t)$ is decomposed into $K$ commuting groups evaluated at discrete times $t_j = j \Delta t$, $\Delta t = t/n$ is the time step, and $n$ is the number of Trotter steps. For the XYZ Heisenberg model, the Hamiltonian is naturally decomposed into groups of commuting Pauli terms, where each group can be implemented as a sequence of two-qubit gates (e.g., $R_{XX}$, $R_{YY}$, $R_{ZZ}$ rotations) and single-qubit $Z$ rotations with time-dependent parameters.

\subsubsection{Energy Expectation Values}

The benchmark measures the energy expectation value of the quantum state at different times:
\begin{equation}
    E(t) = \langle \psi(t) | H | \psi(t) \rangle = \sum_{k=1}^{K} \langle \psi(t) | H_k | \psi(t) \rangle,
    \label{eq:energy_expectation}
\end{equation}
where each term $\langle \psi(t) | H_k | \psi(t) \rangle$ is computed by measuring the corresponding Pauli string expectation value. The algorithm computes:
\begin{itemize}
    \item \textbf{Initial energy}: $E_{\text{initial}} = E(0) = \langle \psi(0) | H | \psi(0) \rangle$
    \item \textbf{Final energy}: $E_{\text{final}} = E(t) = \langle \psi(t) | H | \psi(t) \rangle$ after $n$ Trotter steps
    \item \textbf{Energy change}: $\Delta E = E_{\text{final}} - E_{\text{initial}}$
\end{itemize}

For the dense state vector backend, the energy is computed efficiently by directly evaluating $\langle \psi | H | \psi \rangle$ using the state vector representation. For the MPS backend, the energy is computed using tensor network contractions that exploit the local structure of the Hamiltonian terms.

\subsubsection{Circuit Construction}

\begin{figure}[t]
    \centering
    \includegraphics[width=\columnwidth]{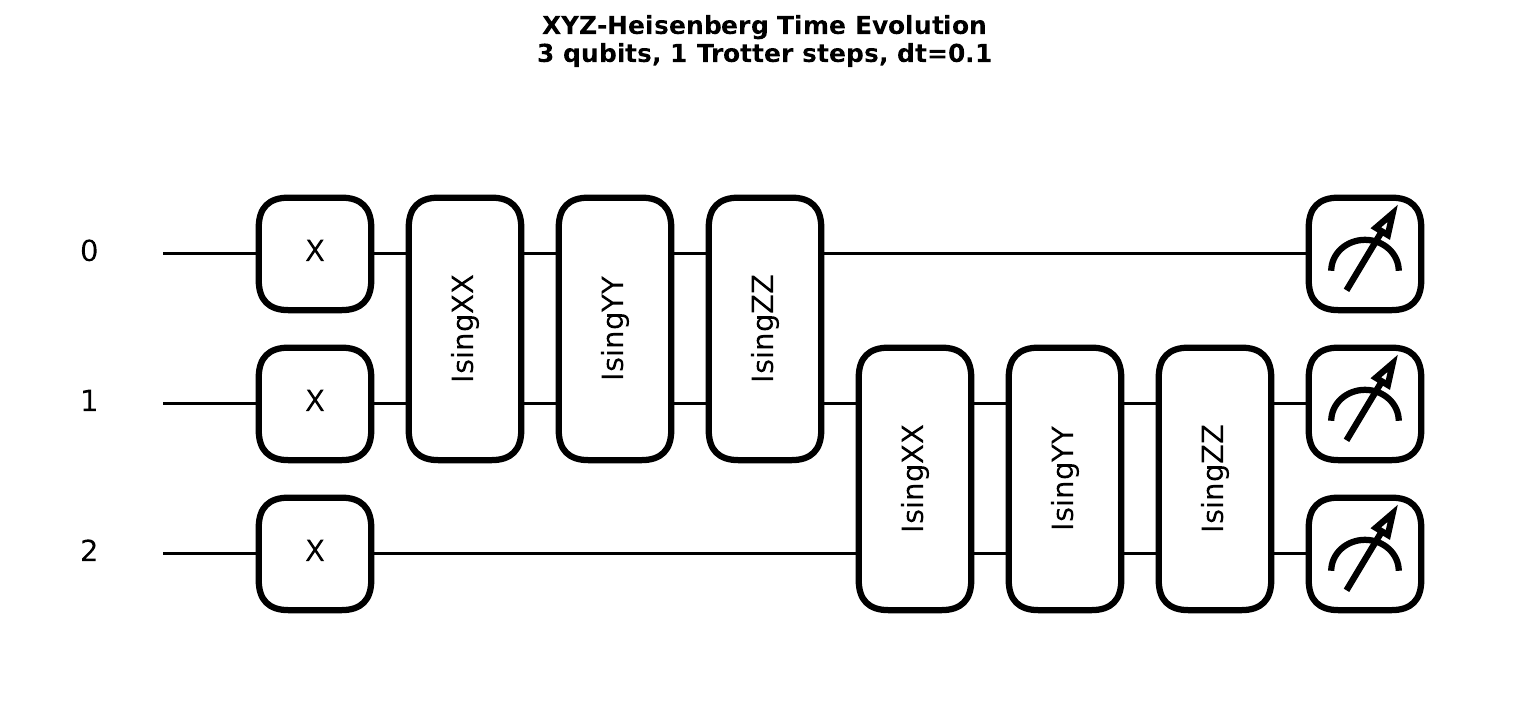}
    \caption{Time evolution circuit for the XYZ Heisenberg model, with decomposition of the Trotter step into a sequence of quantum gates in XX, YY, and ZZ rotations, and single-qubit Z rotations.}
    \label{fig:xyz_h_circuit}
\end{figure}

The time evolution circuit is constructed by decomposing each Trotter step into a sequence of quantum gates. Each term $e^{-iH_k(t_j) \Delta t}$ in the Trotter decomposition is implemented as:
\begin{itemize}
    \item Two-qubit interaction terms: $e^{-iJ_x X_i X_j \Delta t}$ is implemented as $R_{XX}(2J_x \Delta t)$ gates, similarly for $Y_i Y_j$ and $Z_i Z_j$ terms
    \item Single-qubit field terms: $e^{-ih(t_j) Z_i \Delta t}$ is implemented as $R_Z(2h(t_j) \Delta t)$ gates, where $h(t_j) = A \sin(\omega t_j)$ with $A = 2.0$ and $\omega = 1.0$
\end{itemize}
The total number of operations scales as $O(n \cdot N)$, where $n$ is the number of Trotter steps and $N$ is the number of qubits. LogosQ's compile-time type system ensures correct parameter identification and gate sequence construction during circuit generation, including time-dependent parameters.

\subsubsection{MPS Time Evolution Implementation}

\begin{figure*}[t]
    \centering
    \includegraphics[width=0.48\textwidth]{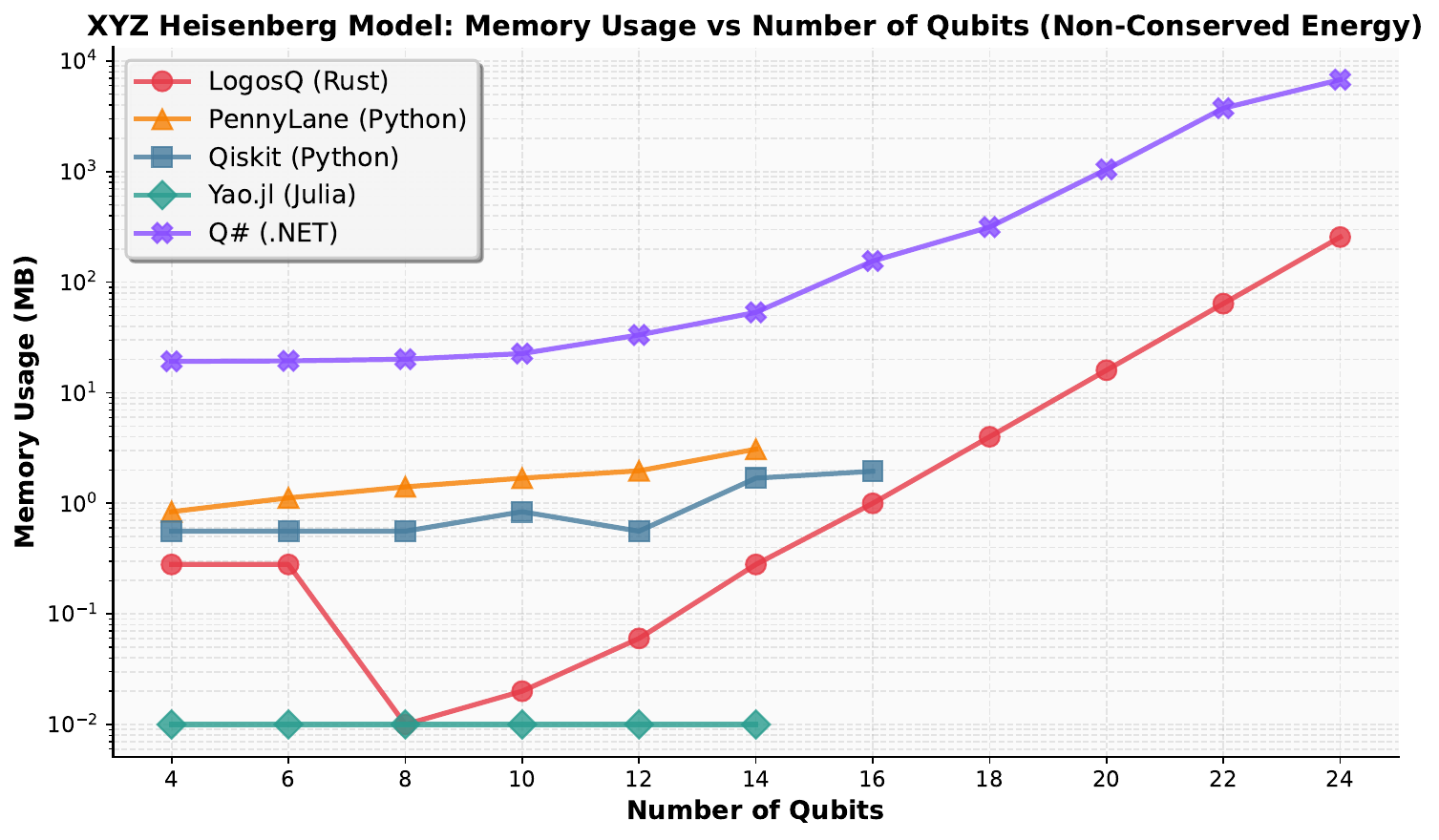}
    \includegraphics[width=0.48\textwidth]{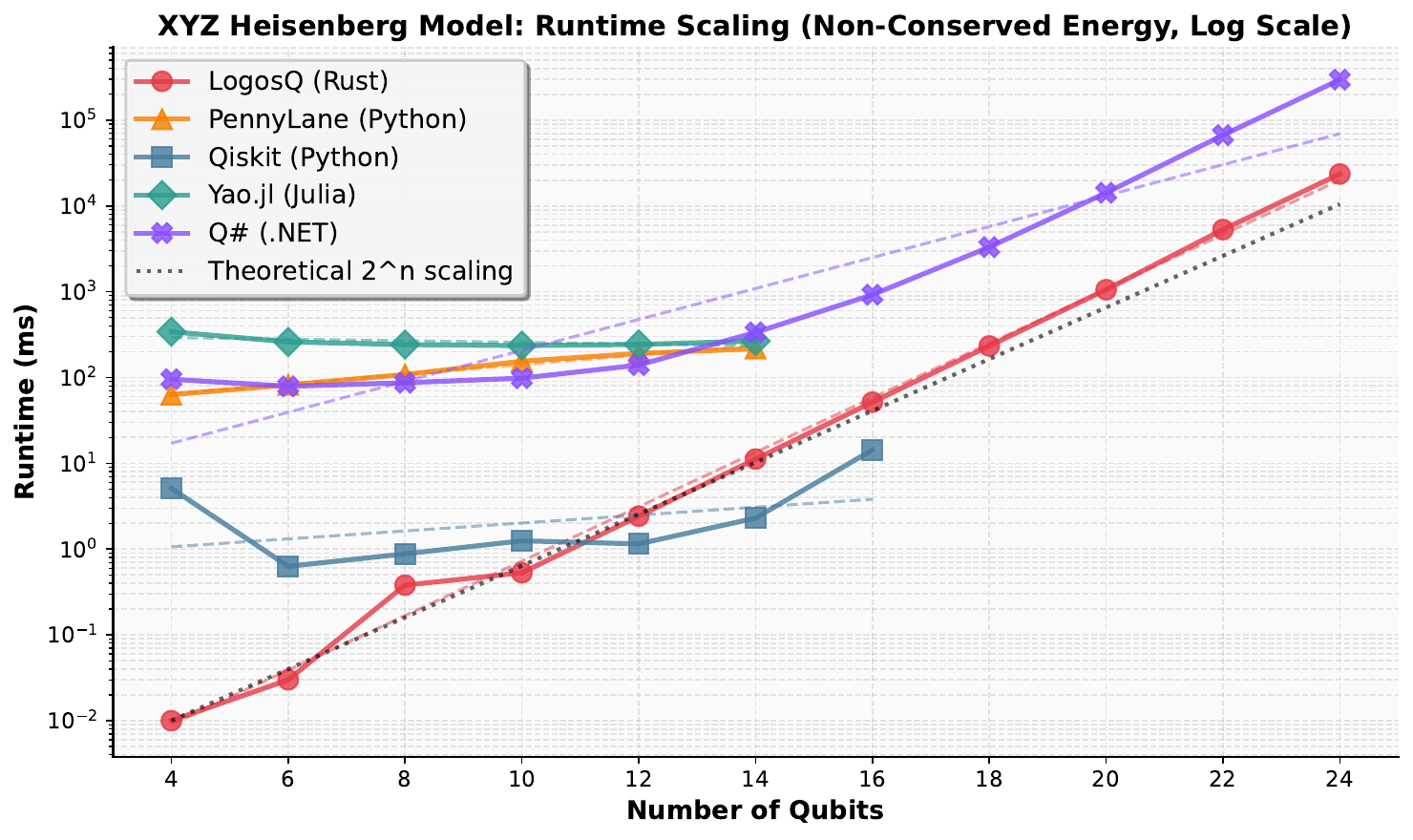}
    \caption{Performance comparison for the XYZ Heisenberg benchmark across LogosQ, PennyLane, Qiskit, Yao.jl, and Q\#. (Left) Memory usage comparison showing LogosQ's efficient memory consumption with the MPS backend, maintaining low memory usage up to 24--25 qubits. (Right) Runtime scaling comparison on a logarithmic scale as the number of qubits increases from 4 to 25. LogosQ sustains a lower slope due to aggressive circuit simplifications and efficient MPS tensor operations. The adaptive backend selection enables simulation of systems that would be intractable with full state vector methods.}
    \label{fig:xyz_performance}
\end{figure*}

For large systems, LogosQ implements time evolution directly on the MPS representation\footnote{\url{https://github.com/zazabap/LogosQ/blob/main/src/simulators/mps.rs}}, using the Time-Evolving Block Decimation (TEBD) approach \cite{vidal_efficient_2003} to apply Trotter gates directly to the tensor network. The detailed mathematical formulation of MPS gate operations, including tensor contraction, SVD decomposition, and truncation procedures, is described in Section~\ref{sec:architecture}. 

For the XYZ Heisenberg model, the MPS backend enables simulation of systems up to 24--25 qubits in our benchmarks while maintaining low memory usage through bond dimensions $\chi \leq 32$ and adaptive truncation. The truncation threshold $\epsilon$ and maximum bond dimension $\chi_{\max}$ are configurable parameters that balance accuracy and computational cost.

\subsection{Adaptive Backend Selection}

For small systems (up to 12--15 qubits), LogosQ employs the state vector backend with $O(2^n)$ memory complexity using direct state vector manipulation (Section~\ref{sec:optimization}). Beyond this limit, LogosQ automatically switches to the matrix product state (MPS) backend (Section~\ref{sec:architecture}). As discussed in Section~\ref{sec:architecture}, the MPS backend reduces memory complexity from $O(2^n)$ to $O(n \chi^2)$ by exploiting the bounded entanglement entropy of one-dimensional systems \cite{cirac_renormalization_2009,schollwoeck_density-matrix_2011}, enabling scalable simulation where bond dimensions typically remain small ($\chi \leq 32$).

\subsection{Benchmark Results}

We benchmark systems ranging from 4 to 25 qubits with isotropic coupling $J_x = J_y = J_z = 1.0$, using identical parameters across frameworks (LogosQ, PennyLane \cite{bergholm_pennylane_2022}, Qiskit \cite{javadi-abhari_quantum_2024}, Yao.jl \cite{luo_yaojl_2020}, and Q\# \cite{svore_q_2018}). All benchmarks were executed on the hardware platform described in Section~\ref{sec:introduction}. For the XYZ-Heisenberg model, PennyLane's maximum simulation capacity on this platform is approximately 14--15 qubits due to memory constraints, while Qiskit and Yao.jl are limited to similar qubit counts. Q\# and LogosQ demonstrate superior scalability, with LogosQ's MPS backend successfully handling systems up to 24--25 qubits with low memory usage across all benchmark tests. While PennyLane-lightning with GPU backend could potentially enhance simulation capabilities, such optimizations are beyond the scope of this discussion.

Figure~\ref{fig:xyz_performance} provides a comprehensive comparison across LogosQ, PennyLane, Qiskit, Yao.jl, and Q\#. LogosQ consistently outperforms other frameworks across all system sizes, with the performance advantage becoming more pronounced as the qubit count increases.
\section{Conclusion}
\label{sec:conclusion}

We have presented LogosQ, a high-performance quantum computing library implemented in Rust that addresses key challenges in quantum software development through compile-time type safety and performance optimizations. LogosQ leverages Rust's type system to provide static verification of quantum circuit correctness, eliminating runtime errors in parameter-shift rule gradient computation for variational algorithms. The library employs direct state vector manipulation, FFT-optimized Quantum Fourier Transform, and adaptive backend selection between state vector and matrix product state (MPS) representations for scalable simulation.

Our benchmarks demonstrate significant performance improvements: speedups of 120--900$\times$ over Python-based frameworks (PennyLane and Qiskit), 6--22$\times$ over Julia-based implementations (Yao.jl), and competitive performance with Q\# \cite{svore_q_2018}. LogosQ achieves chemical accuracy in variational quantum eigensolver experiments on molecular hydrogen and XYZ Heisenberg models, validating both correctness and numerical stability.

The circuit rebuilding architecture, combined with compile-time type checking, provides stronger correctness guarantees than runtime-based dependency tracking approaches, which is essential for variational quantum algorithms where correct gradient computation is critical for optimization convergence.

While LogosQ demonstrates significant advantages, limitations include the state vector simulator's capacity of approximately 10--12 qubits. Although the MPS backend enables scalable simulation of systems up to 24--25 qubits with low memory usage, the current implementation focuses on classical simulation without comprehensive hardware backend interfaces. Future work will extend the MPS implementation with advanced tensor network methods (TTN, PEPS), develop hardware backend interfaces, implement noise models, and explore GPU acceleration. LogosQ's core contributions---compile-time type safety, high performance, and robust gradient computation---provide a solid foundation for quantum algorithm development and research.

\appendix

\section{CNOT Gate Numerical Example}
\label{app:cnot_example}

This section provides a detailed numerical example of the CNOT gate implementation using direct state vector manipulation, complementing the algorithm description in Section~\ref{sec:optimization}.

\subsection{Numerical Example: CNOT on 2-Qubit System}

Consider a 2-qubit system with initial state $\ket{\psi_0} = \frac{1}{\sqrt{2}}(\ket{00} + \ket{11})$, represented as $\psi_0 = [\frac{1}{\sqrt{2}}, 0, 0, \frac{1}{\sqrt{2}}]^T$. Applying CNOT with control $c=0$ and target $t=1$:
\begin{itemize}
    \item For $i=2$ (binary \texttt{10}): Control is set and target is unset, so we swap $\psi[2]$ with $\psi[2 \oplus 1] = \psi[3]$
    \item The state transformation is given by: $\psi[2] \leftrightarrow \psi[3]$ transforms the state to $\psi' = [\frac{1}{\sqrt{2}}, 0, \frac{1}{\sqrt{2}}, 0]^T = \frac{1}{\sqrt{2}}(\ket{00} + \ket{10})$
\end{itemize}
This demonstrates how CNOT flips the target qubit only when the control is set, achieving the transformation through a single amplitude swap operation rather than matrix multiplication.

\section{Toffoli Gate Implementation with Numerical Example}
\label{app:toffoli}

This section provides a detailed implementation of the Toffoli gate using direct state vector manipulation, extending the bitwise optimization principle demonstrated for CNOT in Section~\ref{sec:optimization} (see also Appendix~\ref{app:cnot_example}) to three-qubit controlled gates.

\subsection{Toffoli Gate Algorithm}

\begin{algorithm}
\caption{Direct State Vector Toffoli}
\label{alg:toffoli-app}
\begin{algorithmic}[1]
\REQUIRE State $\psi \in \mathbb{C}^{2^n}$, controls $c_1, c_2$, target $t$
\ENSURE Modified state $\psi' = \text{Toffoli}_{c_1,c_2,t}(\psi)$
\STATE $n \leftarrow$ number of qubits
\STATE $c_{1,\text{bit}} \leftarrow n - 1 - c_1$ \COMMENT{Convert to bit positions}
\STATE $c_{2,\text{bit}} \leftarrow n - 1 - c_2$
\STATE $t_{\text{bit}} \leftarrow n - 1 - t$
\STATE $c_{1,\text{mask}} \leftarrow 2^{c_{1,\text{bit}}}$
\STATE $c_{2,\text{mask}} \leftarrow 2^{c_{2,\text{bit}}}$
\STATE $t_{\text{mask}} \leftarrow 2^{t_{\text{bit}}}$
\STATE $\text{both}_{\text{mask}} \leftarrow c_{1,\text{mask}} \lor c_{2,\text{mask}}$ \COMMENT{Combined control mask}
\FOR{$i = 0$ \TO $2^n - 1$}
    \IF{$(i \land \text{both}_{\text{mask}}) = \text{both}_{\text{mask}}$ \AND $(i \land t_{\text{mask}}) = 0$}
        \STATE $j \leftarrow i \oplus t_{\text{mask}}$ \COMMENT{Flip target when both controls are set}
        \STATE $\text{swap}(\psi[i], \psi[j])$
    \ENDIF
\ENDFOR
\RETURN $\psi$
\end{algorithmic}
\end{algorithm}

Algorithm~\ref{alg:toffoli-app} extends the CNOT optimization to three-qubit gates. 
The condition $(i \land \text{both}_{\text{mask}}) = \text{both}_{\text{mask}}$ ensures both 
control qubits are set, while $(i \land t_{\text{mask}}) = 0$ ensures the target is 
unset before flipping. This maintains $O(2^n)$ complexity without constructing the 
$2^n \times 2^n$ Toffoli matrix \cite{yusufov_designing_2023}. The bitwise operations 
(AND, OR, XOR) are executed in constant time on modern CPUs, making this approach 
highly efficient \cite{smelyanskiy_qhipster_2016}.

\subsection{Numerical Example: Toffoli on 3-Qubit System}

Consider a 3-qubit system with initial state 
$\ket{\psi_0} = \frac{1}{\sqrt{2}}(\ket{110} + \ket{101})$, represented as 
$\psi_0 = [0, 0, 0, 0, 0, \frac{1}{\sqrt{2}}, \frac{1}{\sqrt{2}}, 0]^T$ where indices 
correspond to $\ket{000}, \ket{001}, \ldots, \ket{111}$. Applying Toffoli with controls 
$c_1=0, c_2=1$ and target $t=2$:

For $n=3$ qubits:
\begin{itemize}
    \item $c_{1,\text{bit}} = 3 - 1 - 0 = 2$, $c_{2,\text{bit}} = 3 - 1 - 1 = 1$, $t_{\text{bit}} = 3 - 1 - 2 = 0$
    \item $c_{1,\text{mask}} = 2^2 = 4$ (binary: \texttt{100}, qubit 0)
    \item $c_{2,\text{mask}} = 2^1 = 2$ (binary: \texttt{010}, qubit 1)
    \item $t_{\text{mask}} = 2^0 = 1$ (binary: \texttt{001}, qubit 2)
    \item $\text{both}_{\text{mask}} = 4 \lor 2 = 6$ (binary: \texttt{110})
\end{itemize}

The algorithm checks each state $i$:
\begin{itemize}
    \item For $i=5$ (binary \texttt{101}): Check $(5 \land 6) = 4 \neq 6$ fails (only control $c_1$ is set, not both)
    \item For $i=6$ (binary \texttt{110}): Both controls are set ($6 \land 6 = 6$) and 
        target is unset ($6 \land 1 = 0$), so we swap $\psi[6]$ with 
        $\psi[6 \oplus 1] = \psi[7]$
    \item The manipulation equation: $\psi[6] \leftrightarrow \psi[7]$ transforms the state
    \item Before: $\psi[6] = \frac{1}{\sqrt{2}}$, $\psi[7] = 0$
    \item After: $\psi[6] = 0$, $\psi[7] = \frac{1}{\sqrt{2}}$
    \item Result: $\psi' = [0, 0, 0, 0, 0, \frac{1}{\sqrt{2}}, 0, \frac{1}{\sqrt{2}}]^T = 
        \frac{1}{\sqrt{2}}(\ket{101} + \ket{111})$
\end{itemize}

This demonstrates how Toffoli flips the target qubit (qubit 2) only when both control 
qubits (qubits 0 and 1) are set, transforming $\ket{110} \rightarrow \ket{111}$ through 
a single amplitude swap operation rather than matrix multiplication.


\section{Matrix Product State Computation: Numerical Example}
\label{app:mps_example}

\begin{figure}[t]
    \centering
    \includegraphics[width=\columnwidth]{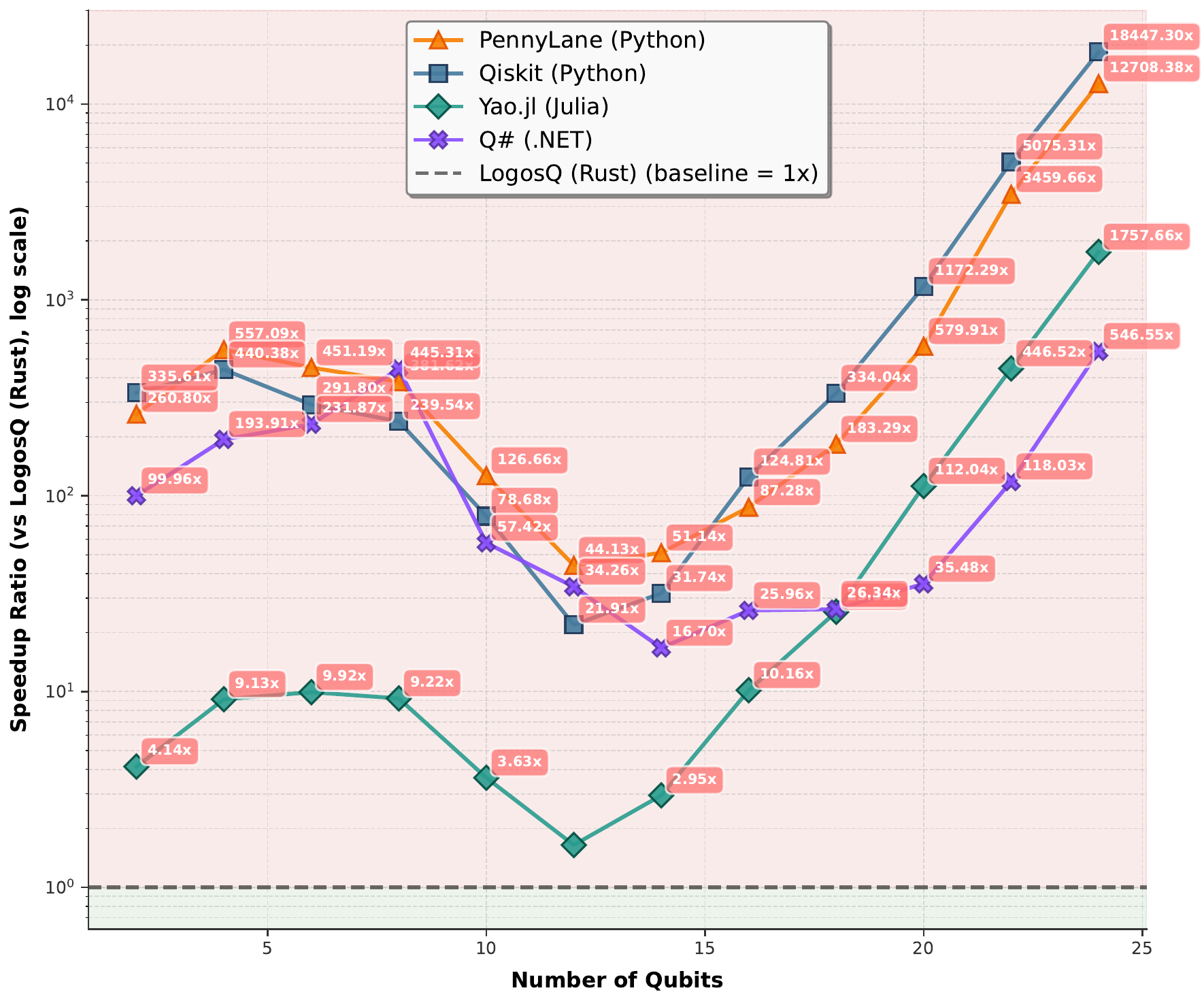}
    \caption{Speedup comparison of QFT implementations across libraries. The MPS-based
    state representation in LogosQ significantly outperforms state-vector-based
    implementations, achieving over $100\times$ speedup at $24$~qubits compared to all
    other libraries in the benchmark.}
    \label{fig:qft_speedup_comparison}
\end{figure}

This section gives a compact, fully worked numerical example of MPS computation.  We
show, step by step, how single- and two-qubit gates are applied to an MPS, and how
entanglement growth appears as an increase in bond dimension, complementing the
theoretical discussion in the MPS section.

\subsection{Initial State Setup}

Consider a 3-qubit system initialized in the computational basis state
\(\ket{\psi_0} = \ket{000}\).  In MPS form, this product state has bond dimension
\(\chi = 1\) (no entanglement).  Each tensor \(A_i\) has shape
\([\chi_{i-1}, 2, \chi_i]\) with \(\chi_0 = \chi_1 = \chi_2 = \chi_3 = 1\).  Explicitly,
\begin{align*}
    A_1^0[1,1] &= 1, & A_1^1[1,1] &= 0 \quad &\text{(shape } [1,2,1]\text{)}, \\
    A_2^0[1,1] &= 1, & A_2^1[1,1] &= 0 \quad &\text{(shape } [1,2,1]\text{)}, \\
    A_3^0[1,1] &= 1, & A_3^1[1,1] &= 0 \quad &\text{(shape } [1,2,1]\text{)}.
\end{align*}
Here \(A_i^s[\alpha_{i-1}, \alpha_i]\) denotes the tensor element with physical index
\(s \in \{0,1\}\) and bond indices \(\alpha_{i-1}, \alpha_i\).

\subsection{Single-Qubit Gate: Hadamard on Qubit 0}

We first apply a Hadamard gate to qubit~0,
\[
    H = \frac{1}{\sqrt{2}}
    \begin{pmatrix}
        1 & 1 \\
        1 & -1
    \end{pmatrix},
\]
using Equation~\ref{eq:mps_single_qubit}.  Only \(A_1\) is affected:
\begin{align*}
    A_1^{0}[1,1] &\leftarrow \sum_{s'_1} H_{0,s'_1} A_1^{s'_1}[1,1]
                 = H_{0,0} \cdot 1 + H_{0,1} \cdot 0
                 = \frac{1}{\sqrt{2}}, \\
    A_1^{1}[1,1] &\leftarrow \sum_{s'_1} H_{1,s'_1} A_1^{s'_1}[1,1]
                 = H_{1,0} \cdot 1 + H_{1,1} \cdot 0
                 = \frac{1}{\sqrt{2}}.
\end{align*}
Thus the updated tensor elements are
\begin{align*}
    A_1^0[1,1] = \frac{1}{\sqrt{2}}, \qquad
    A_1^1[1,1] = \frac{1}{\sqrt{2}},
\end{align*}
corresponding to the state
\[
    \ket{\psi_1}
    = \frac{1}{\sqrt{2}}(\ket{000} + \ket{100}),
\]
with bond dimension still \(\chi = 1\) (no entanglement yet).

\subsection{Two-Qubit Gate: CNOT on Qubits 0 and 1}

Next we apply a CNOT gate with control qubit 0 and target qubit 1 using the
four-step procedure of Section~\ref{sec:mps}.  Because the bond dimension is
currently \(1\), all intermediate tensors are small and can be written as simple
\(2 \times 2\) matrices.

\subsubsection{Step 1: Tensor Contraction}
We first contract \(A_1\) and \(A_2\) along their shared bond
(Equation~\ref{eq:mps_contract}):
\begin{equation*}
    \Theta_{\alpha_0, s_1, s_2, \alpha_2}
    = \sum_{\alpha_1} A_1^{s_1}[\alpha_0,\alpha_1]
                        A_2^{s_2}[\alpha_1,\alpha_2].
\end{equation*}
With \(\alpha_0 = \alpha_1 = \alpha_2 = 1\), the non-zero elements are
\begin{align*}
    \Theta_{1,0,0,1} &= A_1^0[1,1] A_2^0[1,1]
                     = \frac{1}{\sqrt{2}}, \\
    \Theta_{1,1,0,1} &= A_1^1[1,1] A_2^0[1,1]
                     = \frac{1}{\sqrt{2}},
\end{align*}
and all remaining entries vanish.  Grouping indices
\((s_1,s_2)\) into matrix rows and columns, we can write
\begin{equation*}
    \Theta =
    \begin{bmatrix}
        \tfrac{1}{\sqrt{2}} & 0 \\
        \tfrac{1}{\sqrt{2}} & 0
    \end{bmatrix},
\end{equation*}
where rows index \(s_1 \in \{0,1\}\) and columns index \(s_2 \in \{0,1\}\).

\subsubsection{Step 2: Gate Application}
The CNOT operator in the computational basis is
\[
    U_{\text{CNOT}}
    = \begin{pmatrix}
        1 & 0 & 0 & 0 \\
        0 & 1 & 0 & 0 \\
        0 & 0 & 0 & 1 \\
        0 & 0 & 1 & 0
      \end{pmatrix},
\]
as in Equation~\ref{eq:mps_apply_gate}.  Acting on \(\Theta\) produces
\begin{equation*}
    \Theta'_{\alpha_0,s_1,s_2,\alpha_2}
    = \sum_{s'_1,s'_2}
        U_{(s_1,s_2),(s'_1,s'_2)}
        \Theta_{\alpha_0,s'_1,s'_2,\alpha_2}.
\end{equation*}
Because only the components with \(s'_2 = 0\) are non-zero in \(\Theta\),
the action of CNOT simply flips the target qubit when the control is \(1\),
yielding
\begin{equation*}
    \Theta' =
    \begin{bmatrix}
        \tfrac{1}{\sqrt{2}} & 0 \\
        0 & \tfrac{1}{\sqrt{2}}
    \end{bmatrix}.
\end{equation*}

\subsubsection{Step 3: Reshape and SVD}

We now reshape \(\Theta'\) into a \(2 \times 2\) matrix by grouping indices
\((\alpha_0,s_1) \rightarrow \text{row}\) and
\((s_2,\alpha_2) \rightarrow \text{column}\):
\begin{equation*}
    M =
    \begin{bmatrix}
        \tfrac{1}{\sqrt{2}} & 0 \\
        0 & \tfrac{1}{\sqrt{2}}
    \end{bmatrix}.
\end{equation*}
Performing the SVD (Equation~\ref{eq:mps_svd}), \(M = U \Sigma V^\dagger\), we obtain
\begin{align*}
    U &= \begin{bmatrix} 1 & 0 \\ 0 & 1 \end{bmatrix}, \qquad
    \Sigma = \begin{bmatrix}
        \tfrac{1}{\sqrt{2}} & 0 \\
        0 & \tfrac{1}{\sqrt{2}}
    \end{bmatrix}, \qquad
    V = \begin{bmatrix} 1 & 0 \\ 0 & 1 \end{bmatrix},
\end{align*}
with singular values \(\sigma_1 = \sigma_2 = 1/\sqrt{2}\).

\subsubsection{Step 4: Truncation and Canonical Form}

With \(\chi_{\max} = 2\) and truncation threshold \(\epsilon = 10^{-12}\), both
singular values are retained.  To restore canonical form, we absorb \(\Sigma\) into
the left tensor and reshape back into MPS tensors,
\begin{align*}
    A_1^0[1,1] &= \frac{1}{\sqrt{2}}, & A_1^0[1,2] &= 0, \\
    A_1^1[1,1] &= 0,                    & A_1^1[1,2] &= \frac{1}{\sqrt{2}},
\end{align*}
so that \(A_1\) now has shape \([1,2,2]\) (bond dimension increased from \(1\) to
\(2\)).  The right tensor \(A_2\) is updated from \(V^\dagger\) to
\begin{align*}
    A_2^0[1,1] &= 1, & A_2^0[2,1] &= 0, \\
    A_2^1[1,1] &= 0, & A_2^1[2,1] &= 1,
\end{align*}
with shape \([2,2,1]\).  The bond dimension between sites 1 and 2 has therefore
increased from \(\chi = 1\) to \(\chi = 2\), directly reflecting the entanglement
created by CNOT.

\subsection{Resulting State}

The final state is $\ket{\psi_2} = \frac{1}{\sqrt{2}}(\ket{000} + \ket{111})$, a GHZ 
state. The MPS representation now has bond dimension $\chi = 2$ between qubits 0 and 1, 
and $\chi = 1$ between qubits 1 and 2. This example demonstrates how MPS efficiently 
represents entangled states while maintaining computational tractability through 
controlled bond dimension growth.

\section{Edge Cases Handling in Parameter-Shift Rule Gradient Computation}
\label{app:edge_cases}

\begin{table*}[t]
    \centering
    \resizebox{0.9\textwidth}{!}{%
    \tiny
    \setlength{\tabcolsep}{2pt}
    \begin{tabular}{@{}p{2.0cm}p{3.0cm}p{3.5cm}p{3.5cm}@{}}
    \toprule
    \textbf{Metric / Test} & \textbf{PennyLane (PSR)} & \textbf{Yao.jl (Zygote)} & \textbf{LogosQ (PSR)} \\
    \midrule
    \multicolumn{4}{c}{\textbf{System Properties}} \\
    \midrule
    Differentiation & PSR & Automatic diff. & PSR \\
    Parameter Detection & \textcolor{red}{Failed} & \textcolor{green!60!black}{Success} & \textcolor{green!60!black}{Success} \\
    NaN Detection & N/A (empty arrays) & \textcolor{green!60!black}{No NaN found} & \textcolor{green!60!black}{No NaN found} \\
    Edge Handling & \textcolor{red}{Failed} & \textcolor{green!60!black}{Robust} & \textcolor{green!60!black}{Robust} \\
    Error Reporting & \textcolor{red}{Silent failure} & \textcolor{green!60!black}{Explicit} & \textcolor{green!60!black}{Explicit} \\
    Type Safety & \textcolor{red}{Runtime checks} & \textcolor{red}{Runtime checks} & \textcolor{green!60!black}{Compile-time} \\
    \midrule
    \multicolumn{4}{c}{\textbf{Experimental Results (Gradient Output)}} \\
    \midrule
    Normal \newline $[0.5, 0.3, \dots]$ & \textcolor{red}{$[]$} \newline (Fail) & $[-0.4808, -0.0053, \newline -0.0069, -0.0101]$ & $[-0.4808, -0.0053, \newline -0.0069, -0.0101]$ \\
    \midrule
    Large \newline $[10.0, 5.0, \dots]$ & \textcolor{red}{$[]$} \newline (Fail) & $[0.1072, 0.6889, \newline 0.2348, 0.5258]$ & $[0.1072, 0.6889, \newline 0.2348, 0.5258]$ \\
    \midrule
    Near Zero \newline $[10^{-8}, \dots]$ & \textcolor{red}{$[]$} \newline (Fail) & $[-1.0 \times 10^{-8}, 0.0, \newline 0.0, 0.0]$ & $[-1.0 \times 10^{-8}, 0.0, \newline 0.0, 0.0]$ \\
    \midrule
    $\pi/2$ \newline $[\pi/2, \dots]$ & \textcolor{red}{$[]$} \newline (Fail) & $[-0.5000, -0.5000, \newline 0.0, 0.0]$ & $[-0.5000, -0.5000, \newline 0.0, 0.0]$ \\
    \midrule
    $\pi$ \newline $[\pi, \dots]$ & \textcolor{red}{$[]$} \newline (Fail) & $[0.0, 0.0, \newline 0.0, 0.0]$ & $[0.0, 0.0, \newline 0.0, 0.0]$ \\
    \bottomrule
    \end{tabular}%
    }
    \caption{Comprehensive comparison of edge case handling. The table combines qualitative system properties with quantitative experimental results. PennyLane fails to detect parameters (returning empty arrays), while Yao.jl and LogosQ successfully compute valid gradients. LogosQ provides additional robustness through compile-time type safety.}
    \label{tab:edge_case_results}
\end{table*}

This section supplements the compile-time type safety discussion in Section~\ref{sec:compile_time_safety} by examining numerical stability in parameter-shift rule gradient computation. While compile-time type safety ensures correct parameter tracking and prevents runtime errors, numerical stability remains a critical concern when dealing with edge case parameter values. This section demonstrates how LogosQ's design principles extend beyond type safety to provide robust handling of edge cases that can cause silent NaN (Not-a-Number) errors in Parameter-Shift Rule (PSR) gradient computation.

\subsection{Numerical Stability in Parameter-Shift Rule}

While the parameter-shift rule (PSR) formula is mathematically well-defined (see Section~\ref{sec:compile_time_safety}), certain parameter values (e.g., $\theta = \pi/2$, $\theta = \pi$, or values near zero) can cause numerical instabilities that result in NaN gradients. These errors are particularly problematic because they occur silently without raising exceptions, can corrupt entire training runs, waste computational resources, and are difficult to detect in complex variational quantum circuits.

\subsection{Experimental Setup}

The experimental setup consists of a test circuit designed to expose potential NaN-producing scenarios and five test cases that probe different edge case parameter values.

\subsubsection{Test Circuit}

\begin{figure}[t]
\centering
\includegraphics[width=0.8\columnwidth]{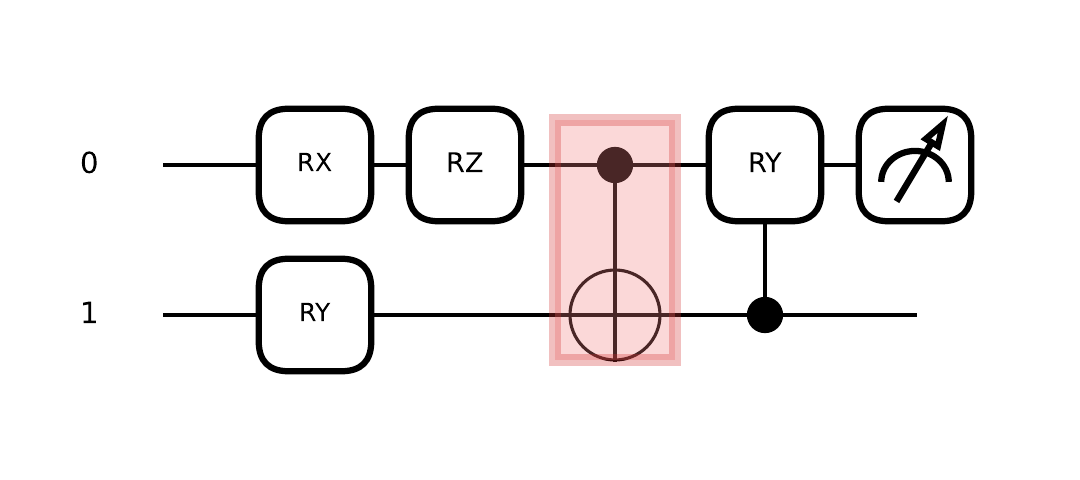}
\caption{Circuit diagram for edge case analysis. The circuit applies $RX(\theta_0)$ on qubit 0, $RY(\theta_1)$ on qubit 1, $RZ(\theta_2)$ on qubit 0, followed by CNOT entangling gate and controlled rotation $CRY(\theta_3)$ where qubit 1 controls rotation on qubit 0. The combination of multiple parameterized gates with a controlled rotation creates scenarios where edge case parameters may cause numerical issues in gradient computation.}
\label{fig:edge_case_circuit}
\end{figure}

The test circuit (Figure~\ref{fig:edge_case_circuit}) uses 4 parameters ($\theta_0, \theta_1, \theta_2, \theta_3$), multiple parameterized gates (RX, RY, RZ rotations), entangling operations (CNOT and controlled rotation CRY), and measures the observable $\langle Z_0 \rangle$ (Pauli Z on qubit 0). This combination creates scenarios where edge case parameters may cause numerical issues in gradient computation.

\subsubsection{Test Scenarios}

Five test cases are evaluated to probe different edge case scenarios:
\begin{itemize}
    \item \textbf{Normal values} $[0.5, 0.3, 0.2, 0.1]$: Baseline behavior
    \item \textbf{Large values} $[10.0, 5.0, 3.0, 2.0]$: Potential overflow or precision loss
    \item \textbf{Near zero} $[10^{-8}, 10^{-7}, 10^{-6}, 10^{-5}]$: Underflow and numerical precision issues
    \item \textbf{At $\pi/2$} $[\pi/2, \pi/2, \pi/2, \pi/2]$: Special angle where trigonometric functions may cause division by zero
    \item \textbf{At $\pi$} $[\pi, \pi, \pi, \pi]$: Another special angle with potential numerical issues
\end{itemize}

\subsection{Experimental Results}

Table~\ref{tab:edge_case_results} presents the gradient computation results for all test cases. PennyLane's implementation shows a \textbf{parameter detection failure} rather than NaN errors, returning empty gradient arrays `[]` for all test cases. This silent failure (no exceptions raised) demonstrates the challenges of runtime parameter dependency tracking when non-parameterized gates are interleaved with parameterized gates, as discussed in Section~\ref{sec:compile_time_safety}.

Yao.jl, using Zygote for automatic differentiation, successfully computes gradients for all edge cases, properly handling special angles ($\pi/2$, $\pi$) and near-zero values without producing NaN or Inf values. LogosQ also successfully computes gradients for all edge cases using the parameter-shift rule, with results matching those of Yao.jl, demonstrating robust numerical stability through its combination of compile-time type safety (Section~\ref{sec:compile_time_safety}) and numerically stable PSR implementation.

\subsection{Analysis of Numerical Stability}

\subsubsection{Sources of Numerical Instability}

Edge cases can legitimately cause NaN errors in PSR implementations through several mechanisms: (1) \textbf{Division by zero} in PSR formulas involving $\sin(\theta)$ or $\cos(\theta)$ in denominators, which can be zero at special angles; (2) \textbf{Precision loss} from near-zero values causing underflow; (3) \textbf{Invalid state computations} when computing $f(\theta \pm s)$ for edge case $\theta$ values; and (4) \textbf{Controlled rotation complexity} where CRY gates may amplify numerical issues.

For a rotation gate $R_G(\theta) = e^{-i\theta G/2}$, the parameter-shift rule requires evaluating $f(\theta + \pi/2) - f(\theta - \pi/2)$. When $\theta = \pi/2$ or $\theta = \pi$, the shifted values become $\theta + \pi/2 = \pi \text{ or } 3\pi/2$ and $\theta - \pi/2 = 0 \text{ or } \pi/2$. These special angles can cause trigonometric functions to evaluate to exact zeros or ones, loss of precision in floating-point arithmetic, and potential division by zero in alternative PSR formulations.

\subsubsection{LogosQ's Robust Edge Case Handling}

LogosQ successfully avoids NaN errors through several design principles that complement its compile-time type safety (detailed in Section~\ref{sec:compile_time_safety}): (1) \textbf{Type-safe parameter tracking} prevents runtime parameter detection failures that could lead to silent errors in edge cases; (2) \textbf{Numerically stable PSR implementation} uses the standard PSR formula that avoids division by trigonometric functions, preventing singularities that can cause NaN; (3) \textbf{Circuit rebuilding architecture} ensures each circuit evaluation is independent, enabling correct gradient computation even for edge case parameter values without state contamination; and (4) \textbf{Explicit error handling} through Rust's memory safety prevents undefined behavior, memory corruption, and implicit type conversions that could cause precision loss or NaN propagation.

This analysis demonstrates that edge case parameter values can cause NaN errors in Parameter-Shift Rule gradient computation. While the specific test cases revealed a parameter detection failure in PennyLane rather than explicit NaN errors, the scenario is theoretically sound and could occur in other implementations or with different circuit configurations. Both Yao.jl (using Zygote automatic differentiation) and LogosQ successfully handle all edge cases, while PennyLane's PSR implementation fails due to parameter detection issues. LogosQ's success demonstrates that its combination of compile-time type safety (Section~\ref{sec:compile_time_safety}) and numerically stable PSR implementation provides robust edge case handling, extending compile-time guarantees beyond preventing type errors to ensuring numerical stability and correctness across a wide range of parameter values.

\printbibliography

\EOD

\end{document}